\documentclass[letter,11pt]{article}
\usepackage{latexsym}
\usepackage{amsmath,amssymb,bm,ascmac,color,calc}
\usepackage{slashed}
\usepackage[mathscr]{eucal}
\usepackage{graphicx}
\usepackage{cite}
\usepackage{lineno}
\usepackage{jheppub}
\usepackage{multirow}
\usepackage{subfigure}
\usepackage[normalem]{ulem}
\usepackage[utf8]{inputenc}
\usepackage[T1]{fontenc}

\usepackage{color}

\usepackage{makecell}
\pdfoutput=1

\allowdisplaybreaks

\newcommand{\met}{{\:/\!\!\!\! E_T}} 
 
\newcommand{\mptvec}{{\;/\!\!\!\! \vec{P}_T}} 
\newcommand{\ignore}[1]{}

\def\beq{\begin{equation}}
\def\eeq{\end{equation}}
\newcommand{\ba}{\begin{array}}
\newcommand{\ea}{\end{array}}
\newcommand{\bea}{\begin{eqnarray}}
\newcommand{\eea}{\end{eqnarray} }
\newcommand{\bal}{\begin{align}}
\newcommand{\eal}{\end{align}}
\def\bi{\begin{itemize}}
\def\ei{\end{itemize}}
\def\ben{\begin{enumerate}}
\def\een{\end{enumerate}}
\def\beq{\begin{equation}}
\def\eeq{\end{equation}}
\def\bc{\begin{center}}
\def\ec{\end{center}}
\def\bt{\begin{table}}
\def\et{\end{table}}
\def\btb{\begin{tabular}}
\def\etb{\end{tabular}}

\def\DeltaR{\Delta R}

\title{Enhanced Higgs pair production from higgsino decay at the HL-LHC}

\author[a,b]{Jianpeng Dai,}
\author[c,b]{Tao Liu,}
\author[a,b]{Daohan Wang,}
\author[a,b]{Jin Min Yang}
 
\affiliation[a]{CAS Key Laboratory of Theoretical Physics, Institute of Theoretical Physics,
Chinese Academy of Sciences, Beijing 100190, P. R. China }
\affiliation[b]{School of Physical Sciences, University of Chinese Academy of Sciences, Beijing 100049, P. R. China}
\affiliation[c]{Institute of High Energy Physics, Chinese Academy of Sciences, Beijing 100049, P. R. China}

\emailAdd{daijianpeng@mail.itp.ac.cn}
\emailAdd{liutao86@ihep.ac.cn}
\emailAdd{wangdaohan@mail.itp.ac.cn} 
\emailAdd{jmyang@itp.ac.cn}
 
\abstract{The scenario of multi-sector SUSY breaking predicts pseudo-goldstinos which are not absorbed by the gravitino and their mass can be as low as ${{\cal O} (0.1)}$ GeV. Since the interactions of pseudo-goldstinos are not so weak as gravitino, a produced higgsino can decay to a pseudo-goldstino plus a Higgs boson insider the detector at the LHC, and thus the higgsino pair production can lead to the signal of Higgs pair plus missing energy. For the scenario of natural SUSY which requires rather light higgsinos, such events may sizably outnumber the Higgs pair events predicted by the SM and be accessible at the HL-LHC  (14 TeV with a luminosity of 3~$\rm{ab}^{-1}$). In this work we examine the observability of such Higgs pair plus missing energy from the decay of light higgsinos produced at the HL-LHC. Considering three channels of the Higgs-pair decay ($bbWW^*$, $bb\gamma\gamma$, $bbbb$), our detailed Monte Carlo simulations for the signal and backgrounds show that the best channel is $bbbb+\met$, whose statistical significance can reach $2\sigma$ level for a light higgsino allowed by current experiments. This is over the SM Higgs pair result which is about $1.8\sigma$.       
}

\begin{document} 
\maketitle

\section{Introduction} \label{sec:intro}

The precision test of the Higgs properties and the exploration of new physics beyond the standard model (SM) is the main task of the Large Ladron Collider (LHC) and other future colliders after the Higgs boson was discovered in 2012. Now the Higgs mass has already been measured to an impressive precision, and in order to confirm the mechanism of electroweak symmetry breaking described in the SM and probe possible new physics beyond the SM, the Higgs self-coupling $\lambda_{hhh}$ has to be reconstructed by experimental measurements which requires the production of at least two Higgs bosons at the LHC. In the SM, the dominant channel for Higgs pair production at the LHC is the gluon fusion channel, in which these two Higgs bosons could be directly radiated from heavy quark loops or   
from a virtual Higgs boson through the self-coupling $\lambda_{hhh}$.
Comparing with the single Higgs production, the Higgs pair processes are
additionally suppressed by the destructive interference between the above two contributions. The cross sections for Higgs pair production is about three orders of magnitude smaller than the single Higgs production, and thus it cannot reach the observable level when considering the huge QCD background at the LHC. On the experimental side, ATLAS and CMS have published measurements of Higgs pair production in the decay channels 
$b{\bar b}WW^*$~\cite{Aaboud:2018zhh, CMS:2017ums},
$b{\bar b}b{\bar b}$~\cite{Aaboud:2018knk, CMS:2018smw},
$b{\bar b}\gamma\gamma$~\cite{Sirunyan:2018iwt, Aaboud:2018ftw},  
$b{\bar b}\tau^+\tau^-$~\cite{Aaboud:2018sfw, Sirunyan:2017djm} and $WW^{*}WW^{*}$~\cite{Aaboud:2018ksn}, and the combination of different measurements gives a constraint on the triliear self-coupling, $-5.0 < \lambda_{hhh}/\lambda_{hhh}^{SM} < 12.0$ at $95\%$ C.L.~\cite{2020cos},
with the assumption that all other couplings are SM-like. The high-luminosity run of the LHC will further narrow the window of $\lambda_{hhh}$. 

The Higgs pair production rate could be altered in new physics models (see, for examples Refs.~\cite{Contino:2010mh, Grober:2010yv, Dolan:2012ac, Cao:2013si, Han:2013sga, Hespel:2014sla, Cao:2014kya, Dawson:2015oha, Lu:2015qqa,
Kang:2015nga, 2016yan, 2017yan, 2017caoyan, Nakamura:2017irk, Chang:2017niy, Ren:2017jbg, Huang:2017nnw, Adhikary:2017jtu, Agrawal:2019bpm, Cheung:2020xij, 2020yan, Abouabid:2021yvw} and
references therein) which may predict different Higgs
self-coupling from the SM, e.g., in the next-to-minimal supersymmetric model (NMSSM) or little Higgs theory, the Higgs self-coupling may be significantly different from the SM value~\cite{Wu:2015nba,Wang:2007zx,Han:2009zp}. Of course, the Higgs pair events may also come from the decays of the new particles predicted in new physics models. In this work we consider supersymmetric models with multi-sector SUSY breaking, in which each breaking sector provides a massless goldstino $\eta_i$ at tree level with SUSY breaking scale $F_i$. When taking into account radiative corrections, the true goldstino which will be absorbed by gravitino remains massless, while the pseudo-goldstino acquires non-vanishing contribution and obtains a mass~\cite{Argurio:2011hs,Dai:2021eah}. Unconstrained by the supercurrent like gravitino, the interaction between pseudo-goldstino and visible fields could be strong enough and thus may lead to unconventional phenomenology~\cite{Cheung:2010mc, Cheung:2010qf, Craig:2010yf, Thaler:2011me, Cheung:2011jq, Argurio:2011gu, Argurio:2011hs, Dudas:2011kt, Liu:2013sx, Hikasa:2014yra, Liu:2014lda, Ferretti:2013wya, Dai:2021eah, Franzosi:2021zwp}. In our previous studies \cite{Hikasa:2014yra,Liu:2014lda,Liu:2013sx} we found that the lightest neutralino could be bino-like and mainly decay to Higgs or longitudinal $Z$ boson plus pseudo-goldstino while the lightest chargino may decay to $W$ boson plus pseudo-goldstino. In those studies only the decay channels to $Z$ or $W$ boson are considered since they are easier to detect.      
In case that the lightest neutralino dominantly decays to Higgs plus pseudo-goldstino, the production of neutralino pair will lead to the signal of Higgs pair plus missing energy.  
Note that as fermions the neutralinos may be more copiously produced than the Higgs bosons. The Higgs pair events from these neutralino decays may sizably outnumber the Higgs pair events predicted by the SM. In addition, the accompanied large missing energy may help to reduce the QCD backgrounds. In this work we perform a comprehensive study for this Higgs channel at the  high luminosity LHC (HL-LHC).   
We will work in the natural SUSY scenario as an example which predicts light higgsinos,  and consider three typical Higgs-pair decay channels to study the observability, i.e., $b{\bar b}WW^*$ with $W$ boson decaying leptonically, $b{\bar b}b{\bar b}$ and $b{\bar b}\gamma\gamma$.      

This work is organized as follows. In Sect.~\ref{sec:II} we make a brief review on pseudo-goldstino and mention some related theoretical basics. Then in Sect.~\ref{sec:III} we perform a Monte Carlo simulation for the signal and backgrounds for three decay channels for the Higgs pair. The conclusion is made in Sect.~\ref{sec:IV}. 

\section{A description of theoretical framework} \label{sec:II}
In the scenario of multi-sector SUSY breaking, 
each hidden sector with spontaneous SUSY breaking at scale $F_i$ could be parameterized in a non-linear way 
$X_i = \eta_i^2/(2F_i) + \sqrt{2} \theta\eta_i + \theta^2F_i$,
where $\eta_i$ denotes the so-called goldstino. Then the soft terms for the visible superfields can be obtained through the non-trivial K\"ahler potential $K$ and gauge kinetic function $f$   
\begin{eqnarray}
K&=&\Phi^\dagger \Phi  \sum_i\frac{ m_{\phi,i }^2}{ F_i^{2} }X_i^\dagger X_i,\\
f_{ab}&=&\frac{1}{g_a^2}\delta_{ab}\left(1+\sum_i\frac{2m_{a,i}}{F_i}X_i\right).
\end{eqnarray}
Here $m_{\phi,a}$ are the soft masses for the chiral superfields
and gauginos, respectively. Other soft trilinear $A$ terms and bilinear $B_\mu$ which will not be used in the following discussion could also be easily constructed. Take the scenario of two hidden sectors as an example, with the definition $F=\sqrt{F_1^2+F_2^2}$ and $\tan\theta=F_2/F_1$, $G=\eta_1\cos\theta+\eta_2\sin\theta$ would be absorbed by gravitino through the super-Higgs mechanism and $G^\prime=-\eta_1\sin\theta+\eta_2\cos\theta$ is left as the physical pseudo-goldstino. Then one can obtain the interaction between $G,G^\prime$ and visible particles up to order $1/F_i$
\begin{eqnarray}
\mathcal{L}_G&=&\frac{m_\phi^2}{F}G\psi\phi^*
-\frac{i m_a}{\sqrt{2}F}G\sigma^{\mu\nu}\lambda^a F^a_{\mu\nu}+
\frac{m_a}{F}G\lambda^a D^a  ,\\
\mathcal{L}_{G^\prime}&=&\frac{\widetilde{m}_\phi^2}{F}G^\prime\psi\phi^*
-\frac{i\widetilde{m}_a}{\sqrt{2}F}G^\prime\sigma^{\mu\nu}\lambda^a F^a_{\mu\nu}+
\frac{\widetilde{m}_a}{F}G^\prime\lambda^a D^a . 
\end{eqnarray}
Through the definition of $m_{a,\phi}$ and $\widetilde{m}_{a,\phi}$
\begin{eqnarray}
m_{a}=m_{a,1}+m_{a,2}, \quad
\widetilde{m}_{a}=-m_{a,1}\tan\theta+m_{a,2}\cot\theta, \nonumber \\
 m^2_{\phi}=m^2_{\phi,1}+m^2_{\phi,2}, \quad
\widetilde{m}^2_{\phi}=-m^2_{\phi,1} \tan\theta+m^2_{\phi,2} \cot\theta,
\end{eqnarray}
it is easy to see that pseudo-goldstino $G^\prime$ could couple to ordinary fields in a total different way compared with gravitino. Under the condition of approximately vanishing $\widetilde{m}_{a}$, the lightest neutralino could only decay to Higgs or longitudinal $Z$ boson plus $G^\prime$. The details about the derivation as well as the corresponding dynamical realization of this condition from the viewpoint of model-buildings could be found in our previous paper~\cite{Hikasa:2014yra}.

Obviously the mass of $G^\prime$ is a crucial parameter for phenomenological analysis. At tree level $G^\prime$ acquires contribution which is just twice the gravitino mass $m_{3/2}$  via supergravity~\cite{Cheung:2010mc}. The leading order correction without gravitational effect arise at three-loop level. When the two messenger scales are equal, the loop correction is at the order of GeV scale~\cite{Argurio:2011hs}. And through an explicit analytical calculation, we found that this correction could be as low as ${\cal O}(0.1)$ GeV if one messenger mass is 100 times larger than the other~\cite{Dai:2021eah}. In our following Motel Carlo simulations the pseudo-goldstino mass is fixed to a reasonable value of $0.5$ GeV.

Next, let us turn to the visible sector and have a close look at neutralino pair production. The property of the neutralinos depends heavily on the choice of certain parameters in the minimal supersymmetric model (MSSM). Since our aim is to study the Higgs phenomenology arising from pseudo-glodstino, we choose the scenario of natural SUSY \cite{Brust:2011tb,Papucci:2011wy,Hall:2011aa,Feng:2012jfa,Baer:2012up,Tata:2020afe}  which predicts light higgsinos. 
In this scenario the higgsinos are rather light, i.e., $\mu$ is assumed to be 100-300 GeV, while the gauginos are much heavier. So the neutralinos $\tilde{\chi}^0_1$ and $\tilde{\chi}^0_2$ are highly higgsino-like and nearly degenerate in masses. Then at the LHC the productions of any pair of them just give missing energy in the MSSM. In order to detect their productions,  a hard jet radiated from initial partons is usually required and the signal of monojet plus missing energy is searched, which is found to be rather challenging at the LHC~\cite{Han:2013usa} (a global likelihood analysis of the electroweakino sector shows that no range of neutralino or chargino masses can be robustly excluded by current LHC searches~\cite{GAMBIT:2018gjo}). Now with multi-sector SUSY breaking, the higgsino-like $\tilde{\chi}^0_1$ and $\tilde{\chi}^0_2$ could decay to Higgs boson plus pseudo-goldstino, and their pair productions can lead to the signal of Higgs pair plus missing energy. 

 \begin{figure}[t]
     \centering
     \includegraphics[width=6cm]{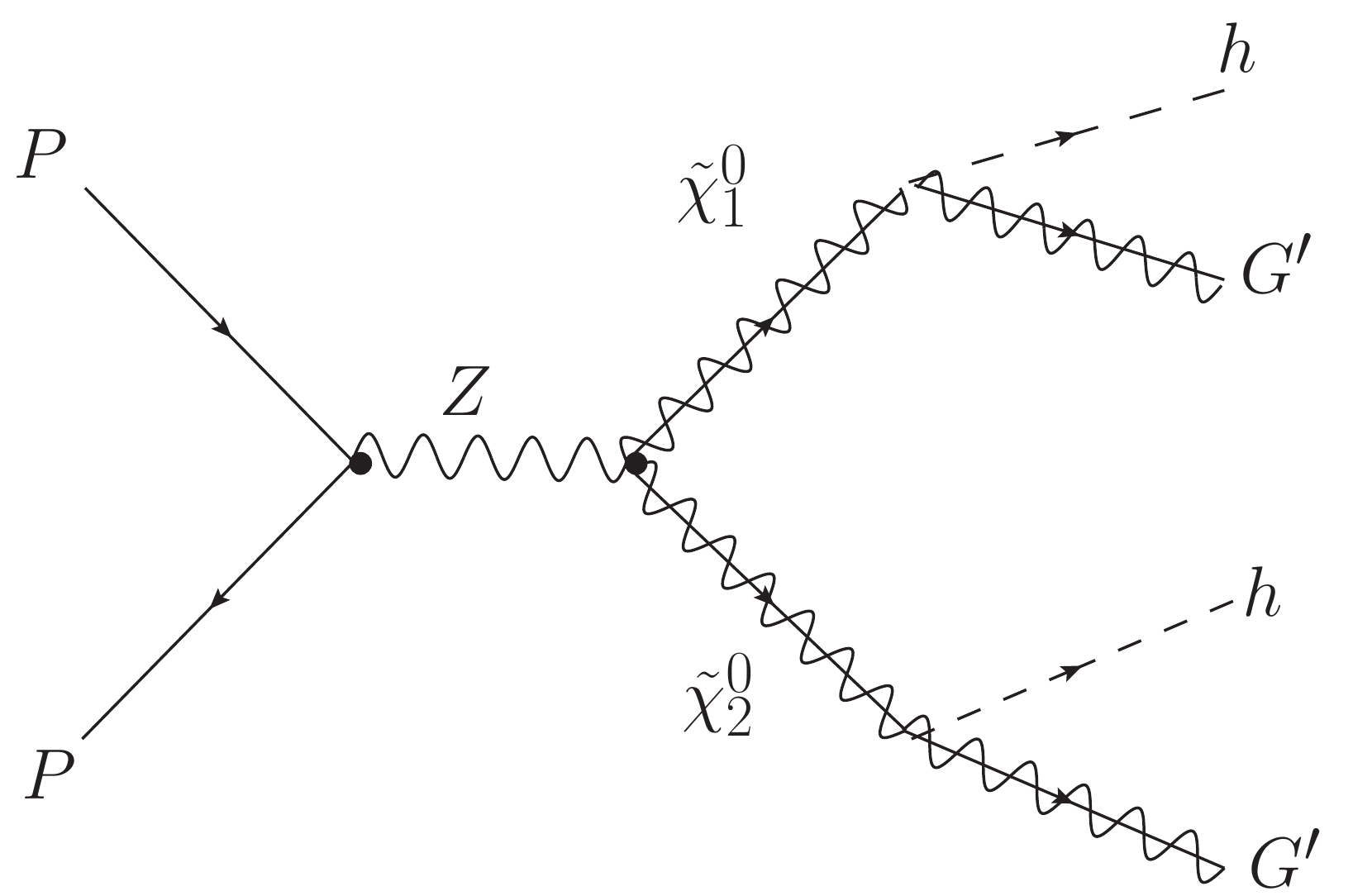}
     \caption{\label{fig:Feynman}
     The Feynman diagram for the Higgs pair event from the decay of higgsinos produced at the LHC }
 \end{figure}
For the pair productions of $\tilde{\chi}^0_1$ and $\tilde{\chi}^0_2$ at the LHC, 
the cross-section of $ \tilde{\chi}^0_1\tilde{\chi}^0_2$ production is much larger than  $\tilde{\chi}^0_1 \tilde{\chi}^0_1 $ and $ \tilde{\chi}^0_2\tilde{\chi}^0_2$.  Thus   
\begin{eqnarray}
 pp \rightarrow \tilde{\chi}^0_1 \tilde{\chi}^0_2 \rightarrow h h G'G'
\end{eqnarray}
is considered in our study. The corresponding Feynman diagram of the signal is shown in Fig.~\ref{fig:Feynman}. To get the mass spectrum and the corresponding mixing matrices for the neutralinos,  SOFTSUSY \cite{2002} is used in our calculation. Here the mass of the SM-liked Higgs boson is fixed at $125$ GeV, $\mu$ is assumed to have a positive sign and the value of $\tan\beta$ is chosen to be 10. In addition, we fix soft gaugino masses as $M_1=1$ TeV and $M_2=1$ TeV.  
  
 \begin{figure}[t]
     \centering
     \includegraphics[width=12cm]{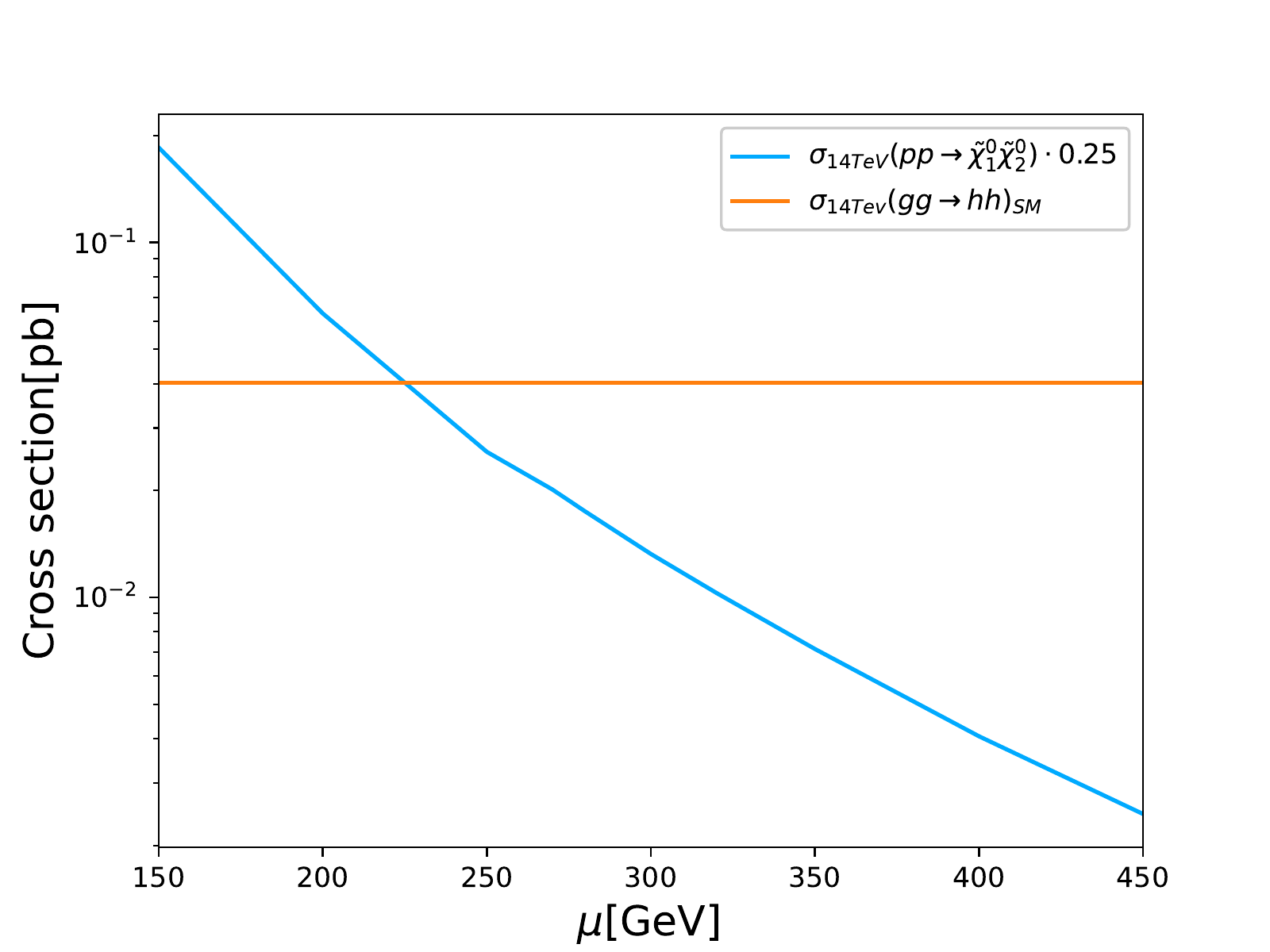}
     \caption{\label{fig:var10} 
     The Higgs pair cross section from $\tilde{\chi}_1^0\tilde{\chi}^0_2$ production at the 14 TeV LHC at next-to-leading order, compared with the SM Higgs pair cross-section at NLO \cite{Borowka:2016ehy,Baglio:2018lrj}. }
 \end{figure}

Before discussing the role of missing energy from pseudo-goldstino in distinguishing signal from backgrounds, in Fig.~\ref{fig:var10} we show the total cross sections for Higgs pair produced through higgsino decay in natural SUSY,  compared with the SM Higgs pair cross-section at NLO~\cite{Borowka:2016ehy,Baglio:2018lrj}. 
 For the couplings between higgsinos and pseudo-goldstino, we use an effective way to study the phenomenology as in the literature~\cite{Hikasa:2014yra} and assume that the branching ratio of higgsino decay to Higgs to be 1/2 throughout this paper. So we see that for light higgsinos the Higgs pair produced from higgsino decay may have a larger cross section than the SM Higgs pair.       
 
\section{Observability at the HL-LHC} \label{sec:III}
In this section we study in detail the signal of Higgs pair plus missing energy from higgsino decay in natural SUSY. Signal and the background processes are modeled using simulated Monte Carlo event samples by MadGraph5\_ aMC@NLO2.4.2~\cite{Alwall:2014hca} with the default NNPDF2.3QED parton distribution functions~\cite{Ball:2013hta} at the $\sqrt{s}=14$ TeV LHC. 
The cross section of higgsino production is normalized to NLO with the help of Prospino2~\cite{Beenakker:1999xh}.
The pseudo-goldstino interaction is implemented in FeynRules~\cite{2014FR} and the UFO model file~\cite{2012UF} is passed to MadGraph5.
We use PYTHIA8.205~\cite{Sjostrand:2014zea} program to describe the parton-shower and hadronization. 

The fast detector simulations are performed by Delphes~\cite{deFavereau:2013fsa} with the ATLAS detector. Using FastJet~\cite{Cacciari:2011ma} for jet-reconstruction with the anti-$k_T$ algorithm~\cite{Cacciari:2008gp}, fixing a cone size of $R$= 0.4 for a jet (the situation of $bbbb$ channel will be further discussed in the following subsection), we include detector effects relevant for the HL-LHC,  where jets
and leptons are smeared according to their energies. For the analysis, we consider jets with $p_{Tj}> 20$ GeV and $|\eta_j|<2.5$, and use the flat $b$-tagging efficiency $\epsilon_{b\rightarrow b} $= 0.75 and the flat mis-tagging rates for non-$b$ jets $\epsilon_{c\rightarrow b}$= 0.1 and $\epsilon_{j\rightarrow b}$= 0.01~\cite{ATL-PHYS-PUB-2019-005}. As for leptons, we require $P^{\ell}_T/\sum P_T>0.7$ with $P_T^{\ell}>20$ GeV and $|\eta_{\ell}|<2.5$ within $\Delta R$=0.3.  

\subsection{The signal of $ h h G'G' \rightarrow b \bar{b} WW^* G'G' \rightarrow  b\bar{b}\ell^+ \ell^- +\met $}
We first study the channel $hhG'G' \rightarrow b \bar{b} WW^* G'G' \rightarrow  bb\ell^+ \ell^- +\met $, which have exactly the same signature as the SM higgs pair production channel $hh \rightarrow b \bar{b} WW^* \rightarrow  bb\ell^+ \ell^- +\met$. The branching ratio of $h\rightarrow WW^*$ is the second largest, next to $h \rightarrow b\bar{b}$. Considering all relevant branching fractions, our signal cross section is $\frac{1}{2}\cdot \sigma_{\tilde{\chi}_1^0 \tilde{\chi}_2^0}\cdot BR(h \rightarrow WW^* \rightarrow \ell^+\ell^-\nu \bar{\nu})\cdot BR(h\rightarrow b\bar{b})$, where $\ell$ denotes an electron or a muon, including leptons from tau decay. Because our signal is similar to SM Higgs pair channels, so we can refer to their backgrounds directly~\cite{2019PDH}. The major background is $t\bar{t}$ production, whose NNLO QCD cross-section is 953.6 pb~\cite{Czakon:2013goa}. The second large background is $t\bar{t}h$, whose NLO QCD cross-section is 611.3 fb~\cite{Dittmaier:2011ti}. For the $t\bar{t}V(V= W^{\pm},Z)$ backgrounds, we apply an NLO k-factor of 1.54 and obtain a cross-section of 1.71 pb~\cite{deFlorian:2016spz}. We also apply a k-factor of 1.0 for the Drell-Yan type background $\tau\tau bb$.  Finally, we generate $tW^{\pm}j$ events (in the five flavor scheme), whose overlap with $t\bar{t}$ should be subtracted.  To reconstruct events, the off-shell effects for the top quark and $W$ boson need to be considered properly.

We employ the following cuts at parton level:
$p_{Tj} > 20$ GeV, 
$p_{Tb} > 20$ GeV, 
$p_{T\gamma} > 10$ GeV, 
$p_{T\ell} > 10$ GeV, 
$|\eta_{j}|$ < 5, 
$|\eta_b|$ < 5, 
$|\eta_{\gamma}|$ < 2.5, 
$|\eta_\ell|$ < 2.5, 
$\Delta R_{bb} < $ 1.8, 
$\Delta R_{\ell\ell} < $ 1.3, 
70 GeV $< m_{jj}, m_{bb}< $ 160 GeV and $m_{\ell\ell} < $ 75 GeV. %
For $tW^{\pm}j$ backgrounds, we impose 5 GeV $< m_{\ell\ell} < $ 75 GeV additionally.
Here the angular distance $\Delta R_{ij}$ is defined by
\begin{eqnarray}
\Delta R_{ij} = \sqrt{(\Delta\phi_{ij})^2+(\Delta \eta_{ij})^2},
\label{deltaR}
\end{eqnarray}
where $\Delta\phi_{ij} = \phi_i-\phi_j$ and $\Delta\eta_{ij}=\eta_i-\eta_j$ are respectively the differences of the azimuthal angles and rapidities between particles $i$ and $j$.
\begin{figure}[t]
     \centering
     \includegraphics[width=7.5cm]{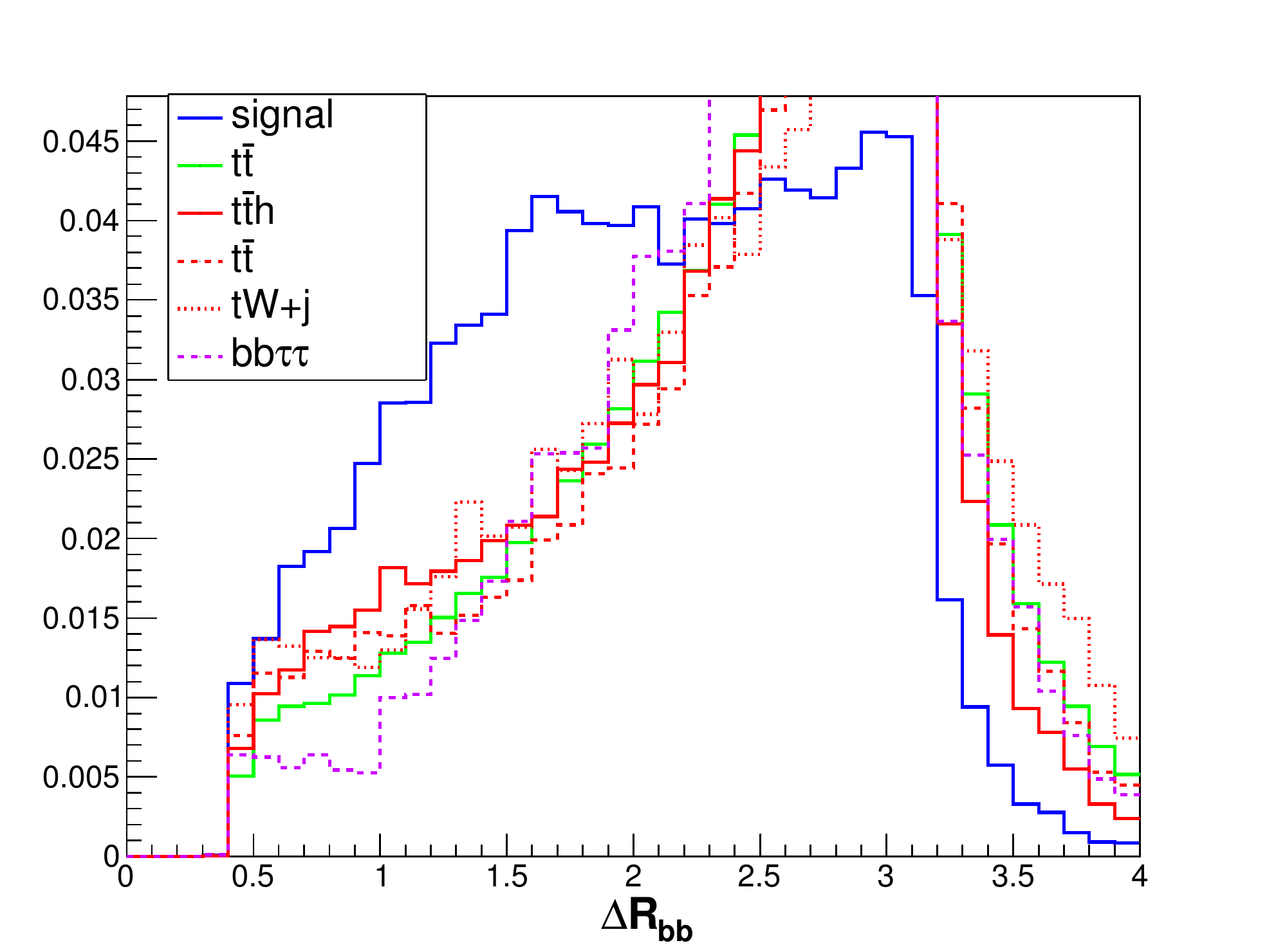}
     \includegraphics[width=7.5cm]{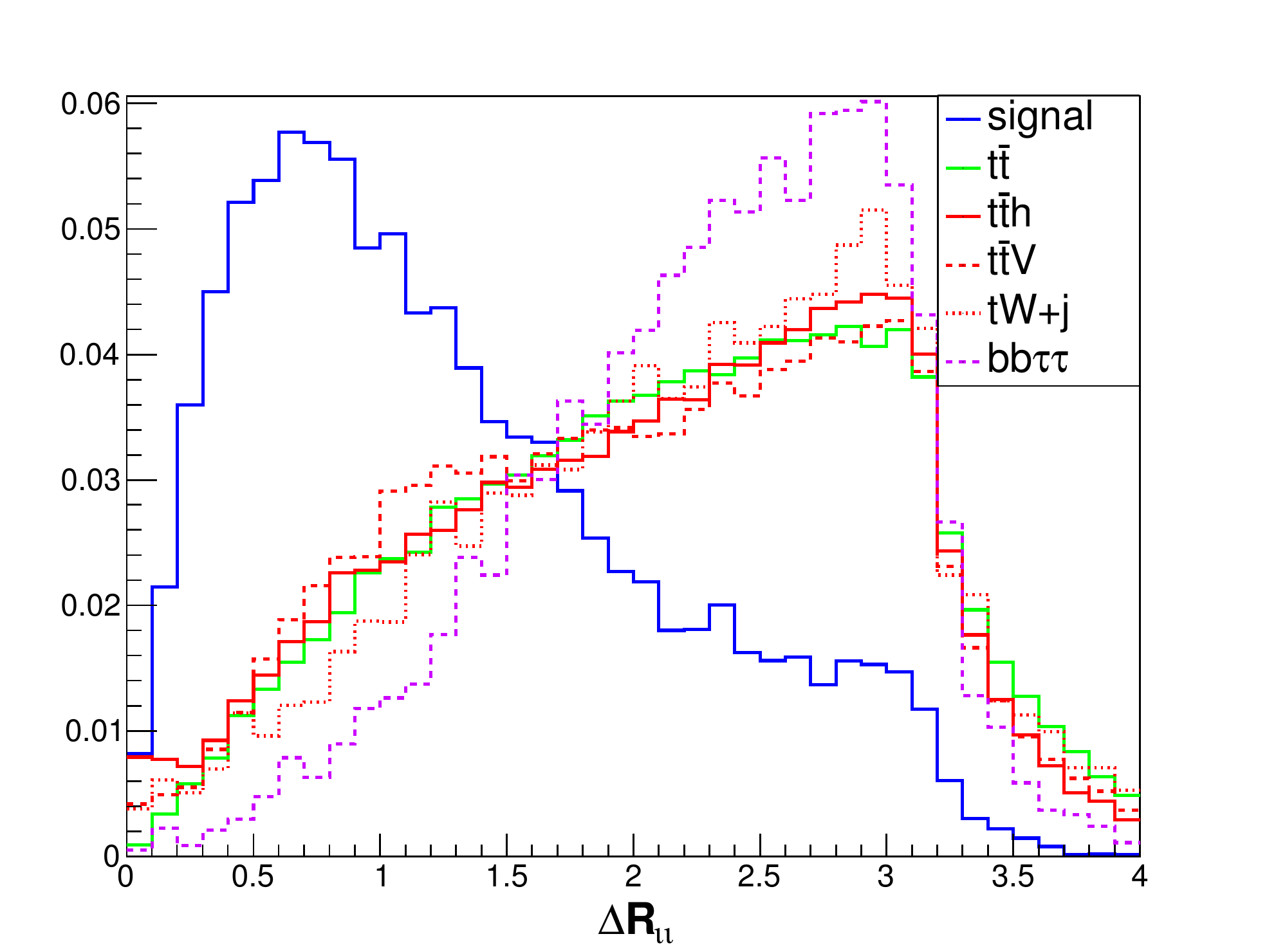}
     \\
     \includegraphics[width=7.5cm]{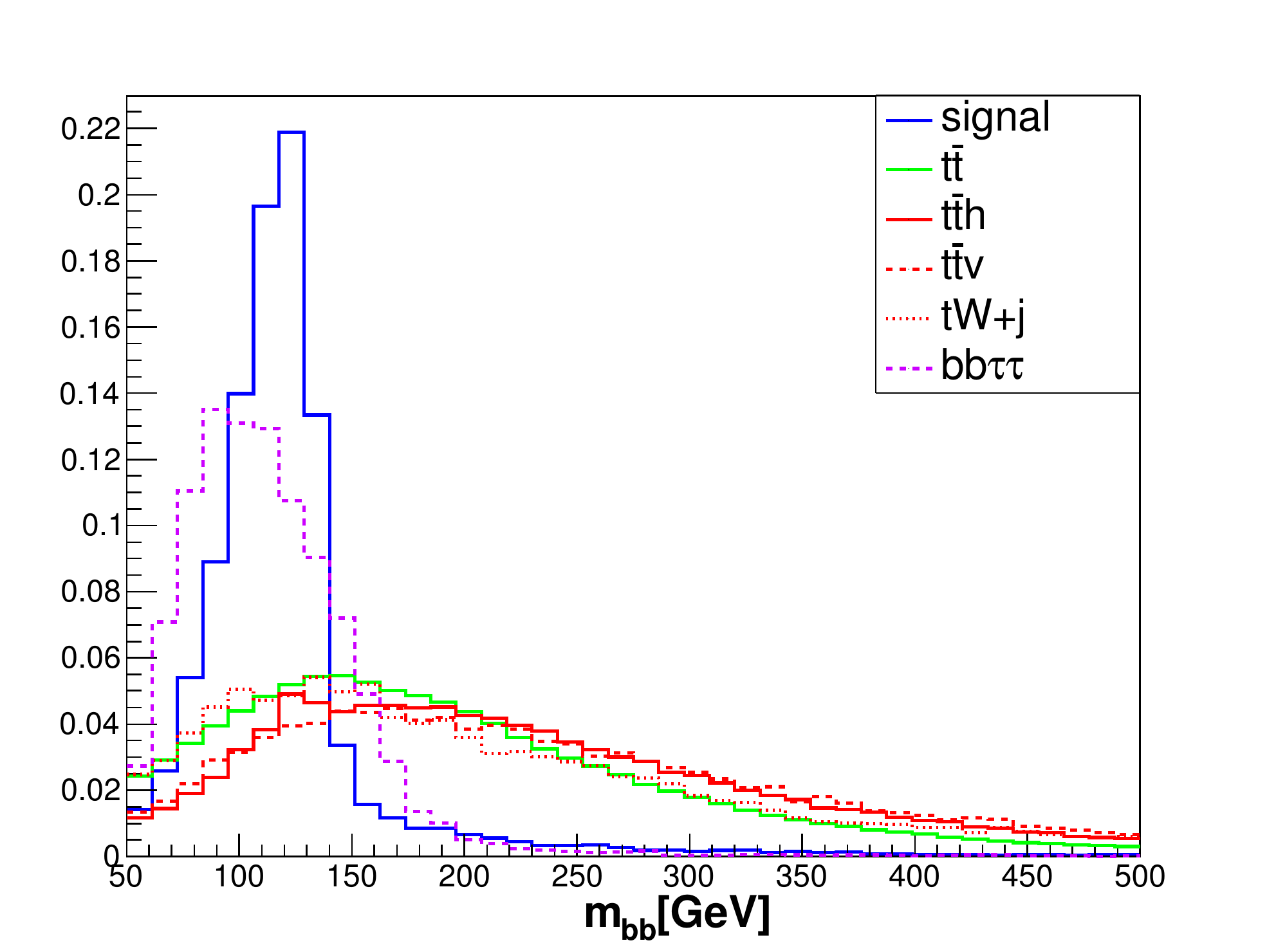}
     \vspace*{0.05cm} 
     \includegraphics[width=7.5cm]{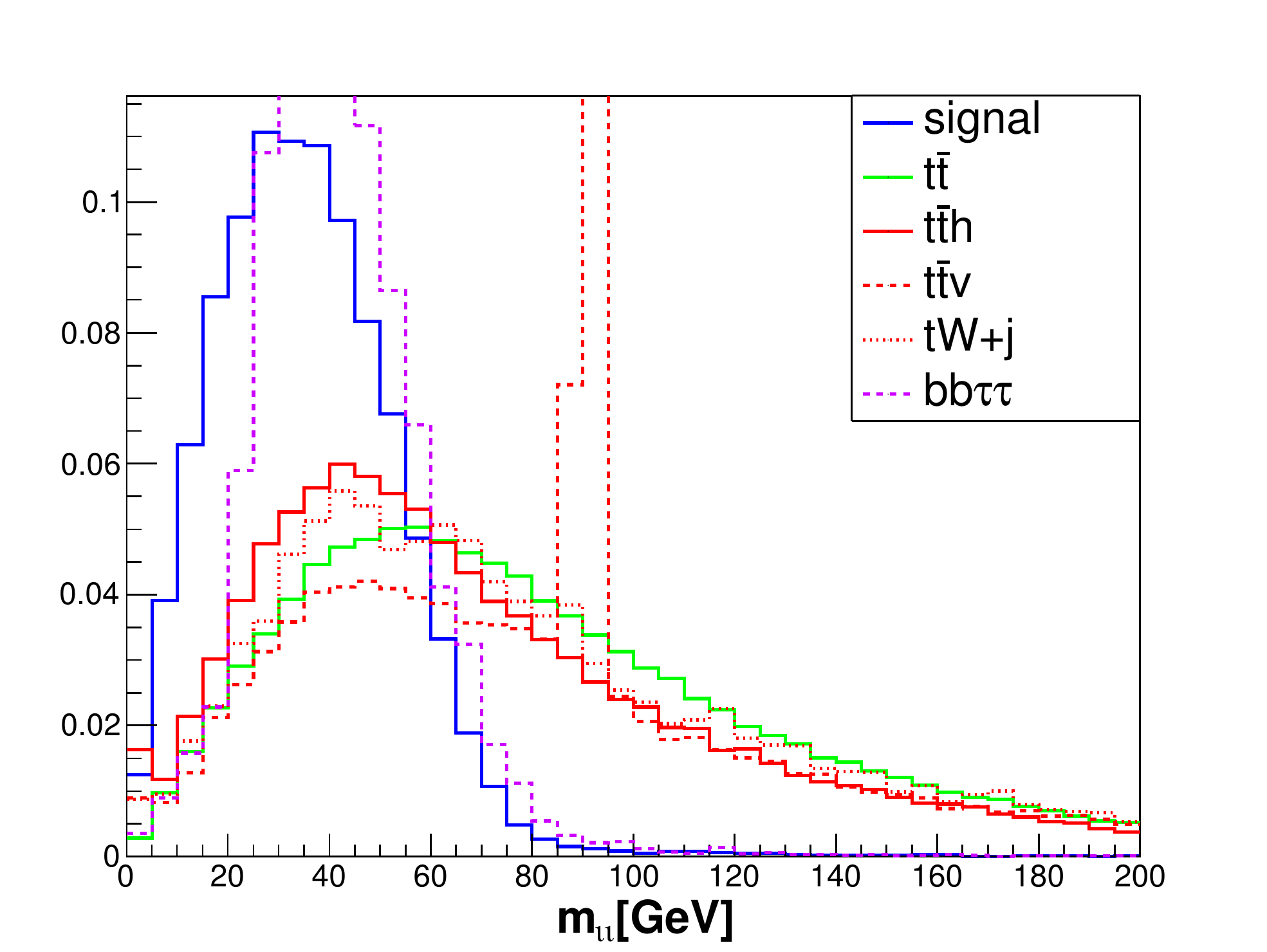}
     \caption{\label{fig:bw}
     Distribution of the $\mu$ =150 GeV for signal and different types of backgrounds after basic cuts. The $y$-axis represents the normalized number of events for each process. }
     
     \hspace{-0.5cm}
 \end{figure}

 We set a sequence of event selections. From Fig.~\ref{fig:bw} we can find that the distributions of $\DeltaR_{\ell\ell}$, $\DeltaR_{bb}$, $m_{bb}$, $m_{\ell \ell}$ and $\met$ are different between the signal and the backgrounds, and thus we use them to make further cuts. Since the $P_T$ distributions of backgrounds and signal have no obvious difference, we simply set $P_{Tl} > 20$ GeV and $P_{Tb}> 30$ GeV.  The cuts $\DeltaR_{ll} < 1.0$ and $\DeltaR_{bb} < 1.0$ are imposed to highly suppress the backgrounds. 
 We make use of invariant masses of leptons and $b-$jets to suppress backgrounds, because the pair of $b-$jets is reconstructed near the Higgs boson mass. From a further analysis of the signal and background features, we find that the missing energy of $t\bar{t}$ is quite approaching to the signal, and thus we gradually increase the cuts of $\met$ to find the best value. Finally, we find that $\met$ > 120 GeV can most efficiently cut the $t\bar{t}$ and $t\bar{t}h$ backgrounds. The detailed results are shown in Table \ref{tab:Cutflowbw}. 
 
\begin{table*}[t]
\caption{
Signal and background cross sections in units of fb after baseline cuts (first row) and at different stages of analysis, using a combination of kinematic variables and requiring the events number $N > 10$. 
The significance $\sigma$ is calculated using the Poisson formula for a luminosity of 3~$\rm{ab}^{-1}$ at the 14 TeV LHC. 
} 
\label{tab:Cutflowbw}
\vspace*{-0.3cm}
\begin{center}
\setlength{\tabcolsep}{1.1mm}
\renewcommand{\arraystretch}{1.4}
\scalebox{0.77}{
\begin{tabular}{|c||c|c|c|c|c|c|c|c|}
\hline   
 Cuts & signal   
& $t\bar{t}$   & $t\bar{t}h$  & $t\bar{t}V$  & $tW+j$  
& $b\bar{b}\tau^+\tau^-$ 
& $\sigma$  & $S/B$
\\
\hline \hline 
$P_{T\ell}$ > 20 GeV, ~~$P_{Tb}$ > 30 GeV
& 0.1008 & 3593.325 & 5.0506
& 10.684 & 144.19 & 1.169
& 0.01 & 2.7$\times10^{-6}$
 \\
\hline
$\DeltaR_{\ell\ell} < 1.0, ~~ \DeltaR_{bb} < 1.0$  & 0.015 & 30.59 & 0.065 & 0.175 & 0.9424 & 0.0254
& 0.146 & 4.7$\times10^{-4}$
\\
\hline
$m_{\ell\ell}$ < 65 GeV, 95 GeV $< m_{bb} < 140$ GeV   & 0.00784 & 0.4746 & 0.0165 & 0.01287 & 0.0228 & 0.0039
& 0.58 & 0.0148
\\
\hline
$\met$ > 50 GeV   & 0.00682 & 0.4  & 0.0144 & 0.0116 & 0.023 & 0.0039
& 0.553 & 0.015
\\
\hline
$\met$ > 80 GeV   & 0.0058 & 0.268 & 0.01 & 0.0099 & 0.0228 & 0.0039
& 0.564 & 0.018
\\
\hline
$\met$ > 120 GeV   & 0.00433 & 0.0603 & 0.006 & 0.0064 & 0.0114 & 0.0039
& 0.793 & 0.049
\\
\hline

\end{tabular}}
\end{center}
\end{table*}

With the above analysis, we can summarize the sequence of cuts as follows:
\begin{itemize}
    \item The two leading jets must be $b-$tagged, each with $P_T$ > 30 GeV;
    \item Exactly two isolated leptons of opposite sign, each with $P_{T\ell}$ > 20 GeV;
    \item Proximity cut of $\DeltaR{_{\ell\ell}}$ < 1.0 for the two leptons;
    \item Proximaty cut of $\DeltaR{_{bb}}$ < 1.0 for two $b-$tagged jets;
    \item $m_{\ell \ell}$ < 65 GeV for two leptons;
    \item 95 GeV < $m_{bb}$ < 140 GeV for the two $b-$tagged jets;
    \item $\met = |\mptvec| > 120$ GeV for the reconstructed missing transverse momentum.
\end{itemize}

From our simulation result, we find the  $bb\tau\tau$ background can be greatly suppressed, while the $t\bar{t}$ background remains to be the dominant one. Finally the significance can only reach 0.79 $\sigma$. Although we can increase the value of $\mu$ to relatively enhance the signal events with $\met > 120$ GeV, this will also suppress the total signal cross section and thus cannot enhance the significance. So we conclude that this channel cannot be observed at the HL-LHC. 

\subsection{The signal of $ hhG'G' \rightarrow bb\bar{b}\bar{b}+\met$}
Now we turn to the decay channel $ hhG'G' \rightarrow bb\bar{b}\bar{b}+\met$.
The decay $h \rightarrow b\bar{b}$ has the largest branching ratio of $58\%$.
In the SM this channel from the Higgs pair production suffers from overwhelming QCD multi-jet backgrounds and its significance can only reach $1.8 \sigma$ at the HL-LHC~\cite{2015NHP}. In the natural SUSY scenario under our consideration, $\met$ of the signal has the same order of magnitude as $\mu$, so the $bbbb$ process could be one of the main backgrounds due to its large production rate, especially when the value of $\mu$ is small.  Another type of dominant backgrounds are the QCD multi-jets plus $W$ or $Z$ boson. In the case of multi-jets plus a $W$ boson,  it can fake the signal when the $W$ boson decays to $\ell \nu$ with the charged lepton missing detection (too soft, travelling along the beamline or too close to a jet) or the $W$ boson decaying to $\tau\nu$ with the secondary jet from the hadronic $\tau$ decay missing detection.  For the case of multi-jets plus a $Z$ boson, it can fake the signal when  $Z$ decays to $\nu \bar{\nu}$. Although their cross sections are about 1-2 pb, mucher larger than the cross section of the signal, they are not the major backgrounds due to the small fake efficiency. The $t\bar{t}h(h\to b\bar{b}),t\bar{t} b\bar{b}$ process, which were not the main background in the preceding section, is now likely the major background. So we consider the backgrounds from $bbbb$, $bbbbZ$, $bbjjW^{\pm}(W^{\pm}\rightarrow \ell \nu)$, $bbjjW^{\pm}(W^{\pm}\rightarrow \tau \nu)$, $bbjjZ$ and $t\bar{t}h(h\to b\bar{b}),t\bar{t} b\bar{b}$ processes.  All these backgrounds are generated using MadGraph with the PDF CTEQ6L1~\cite{2003JPP} at the leading order. Since the cross-section of this signal is large enough even for a relatively large $\mu$, we will choose different values of $\mu$ to show the results. At the parton level,  we simply require $p_{Tj} > 30$ GeV, $|\eta_j|< 2.7$, $\DeltaR_{jj}>0.4$ in our simulations.

\begin{figure}[t]
     \centering
     \includegraphics[width=7.7cm]{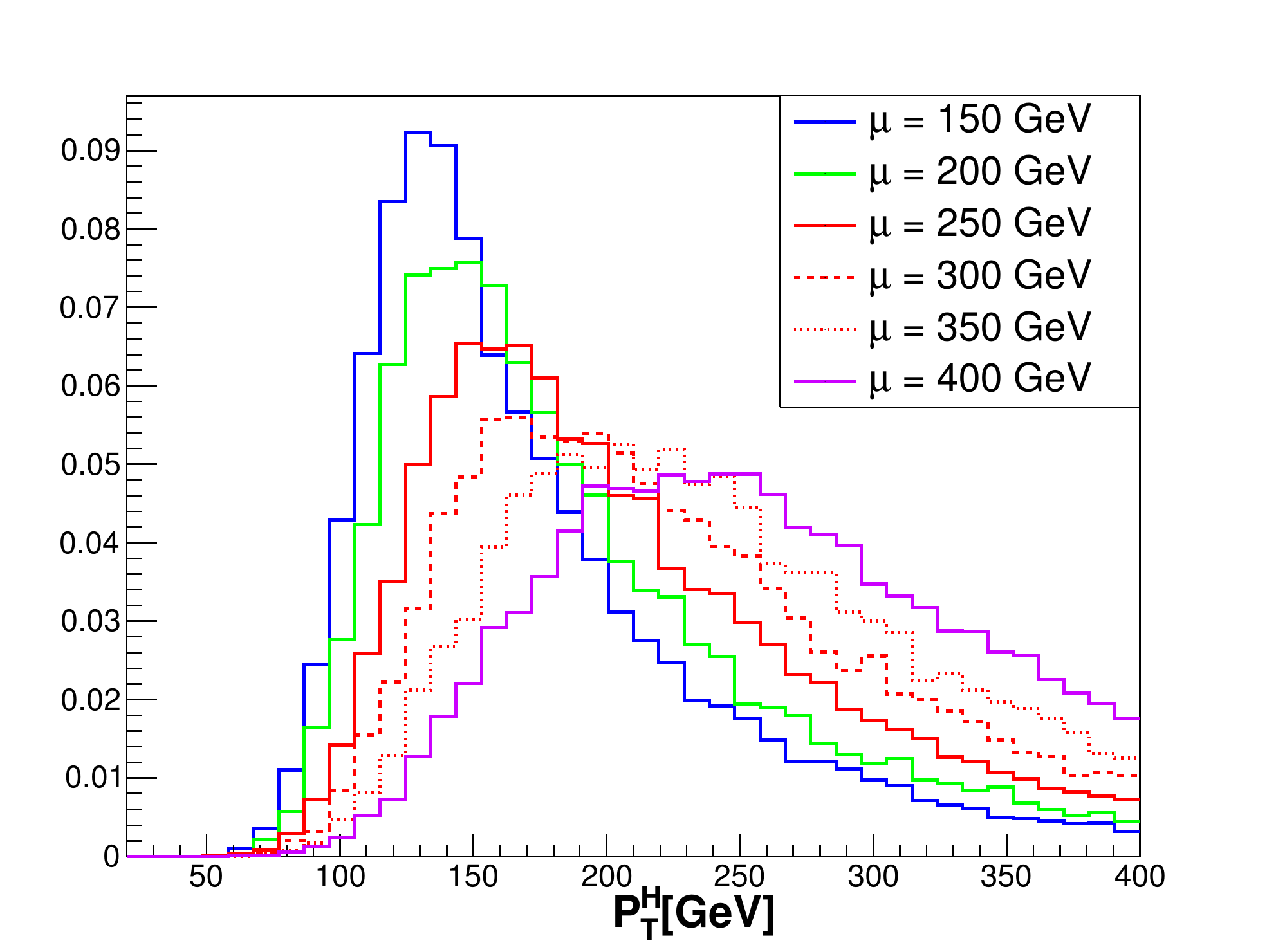}
     \includegraphics[width=7.7cm]{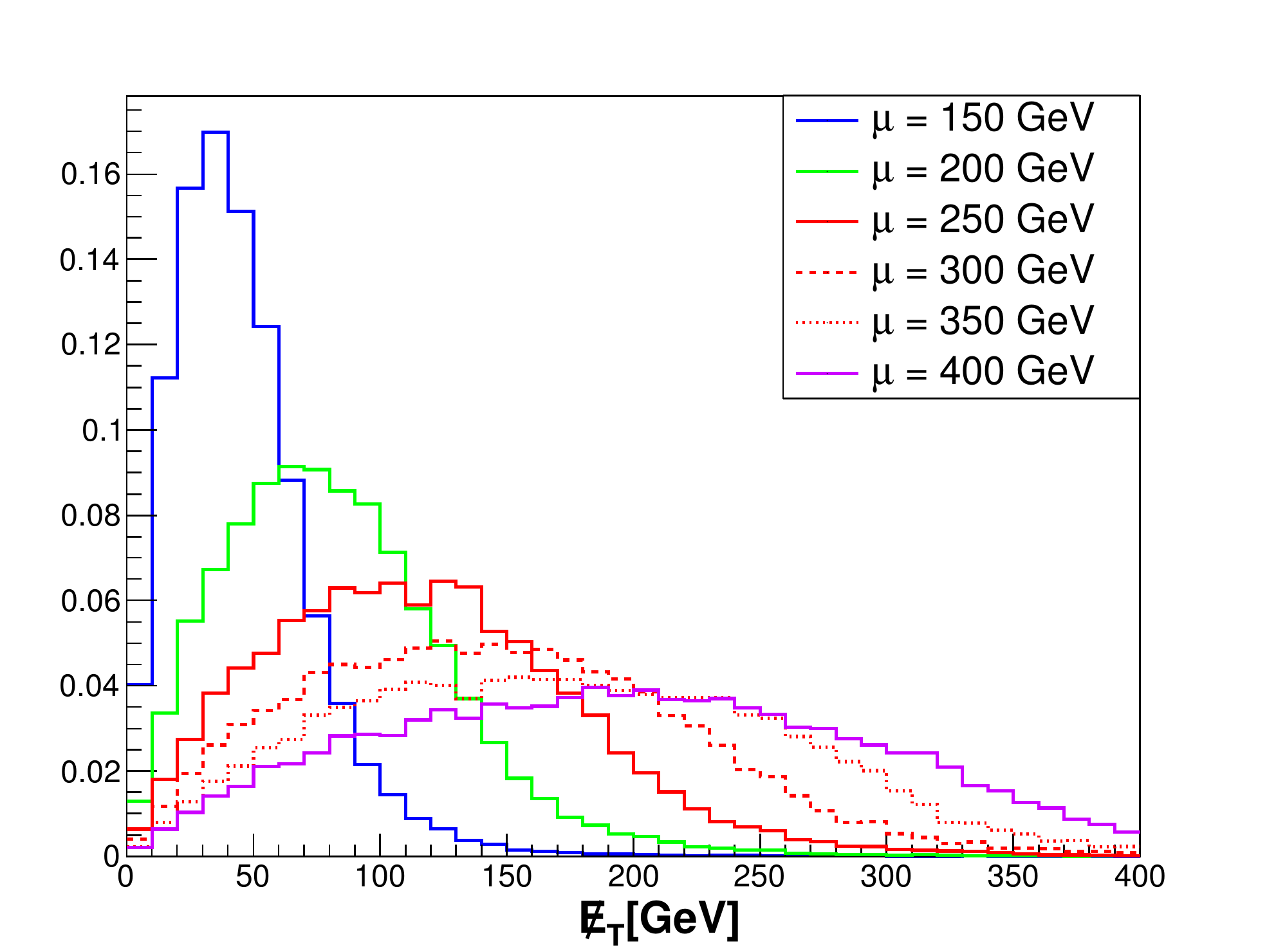}
     \\ \vspace*{-0.1cm}
     \caption{\label{fig:var1b1}
     Distributions of $P_T$ of the leading Higgs and  $\met$  for the signal  $bb\bar{b}\bar{b}+\met$ with different values of $\mu$. The $y$-axis represents the normalized number of events.}
     
     \hspace{-0.5cm}
 \end{figure}

We note that the triggers and jet reconstructions are very important for the signal, in which the transverse momentum of the Higgs boson plays a key role. So after parton-shower and hadronization, we show the $P_{T}$ of the leading Higgs and $\met$ distributions in Fig.~\ref{fig:var1b1}. We find that the signal has relatively low transverse momentum for the Higgs, which is typically below 200 GeV for the value of $\mu$ below 400 GeV. We require at least four anti-$k_t$ R=0.4 $b-$jets for the signal (the efficiency for reconstructing a Higgs boson from two anti-$k_t$ $R=0.4$ jets is higher than reconstructing from a Cambridge-Aachen jet with $R=1.2$ in this case~\cite{2015NHP}). 

So the four $b-$tagged jets are required, paired into two dijets, to reconstruct the Higgs bosons, and this is a powerful way to reduce the backgrounds.Obviously, there are three different ways for each event to pair the b jets. Here the traditional $\chi^2$ method is used to pick out the reconstruction which has the minimum value of $\chi^2$.

\begin{figure}[t]
     \centering
     \includegraphics[width=7.7cm]{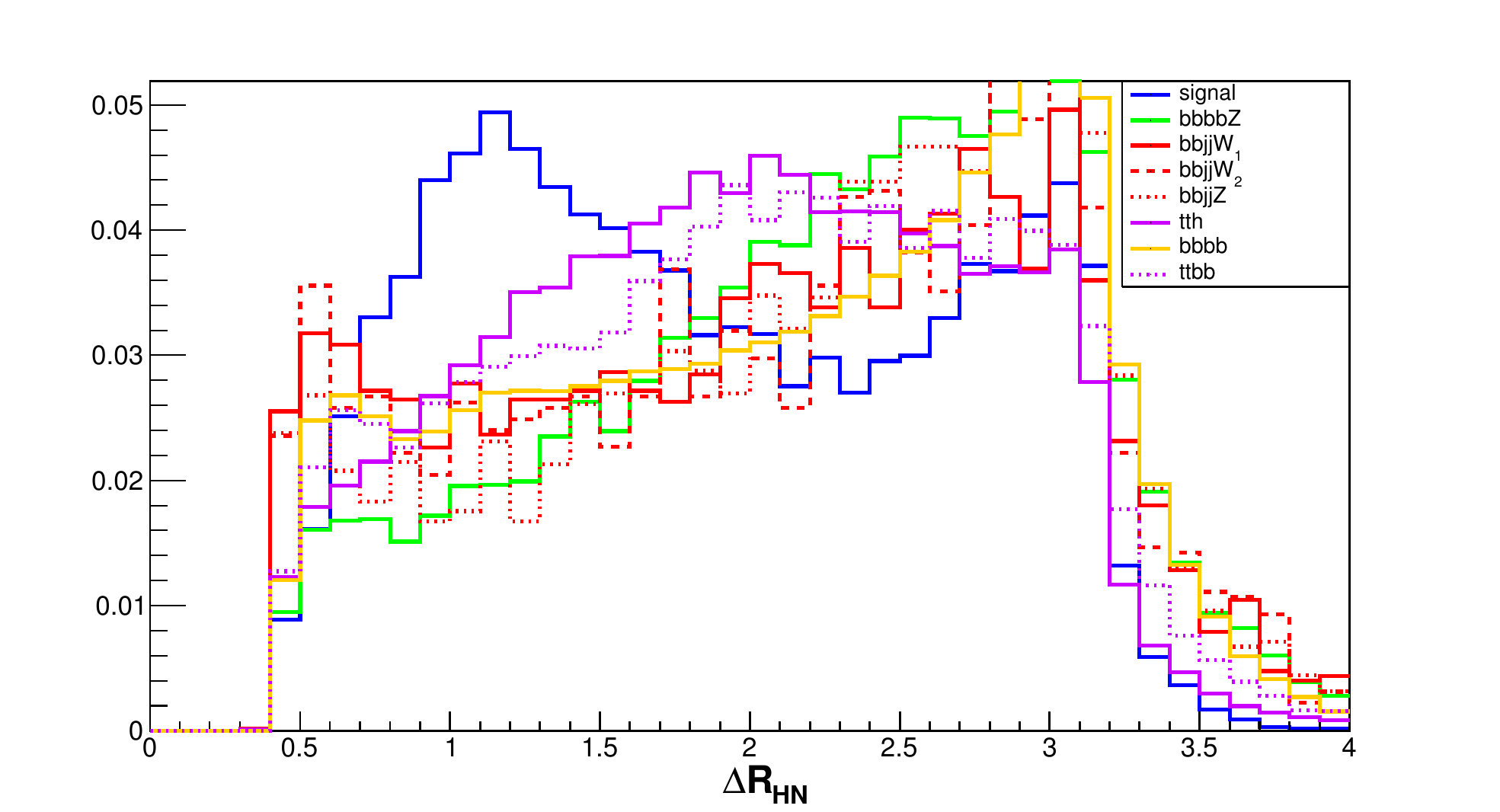}
     \includegraphics[width=7.7cm]{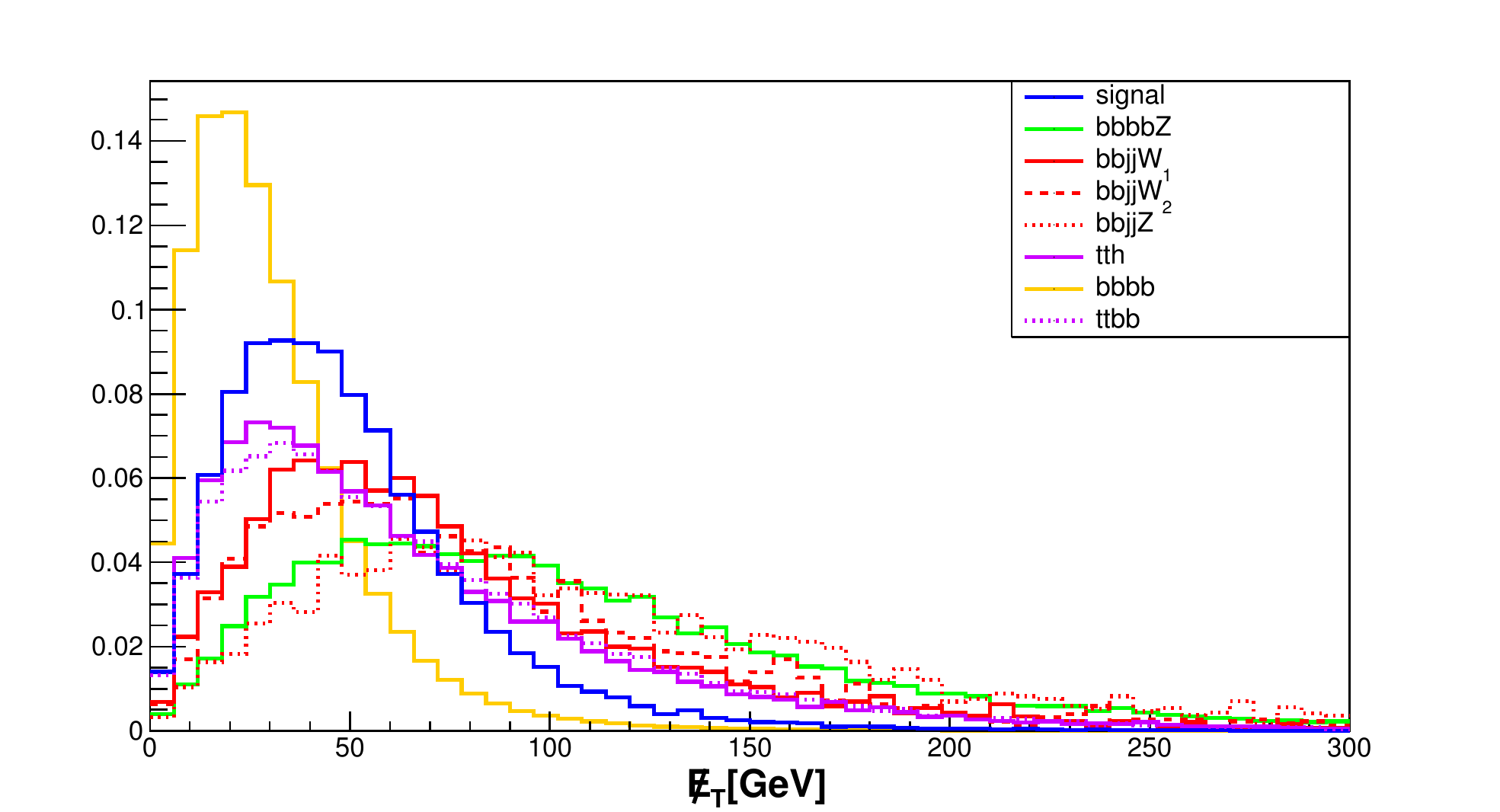}
     \\ \vspace*{-0.1cm}
     \includegraphics[width=7.7cm]{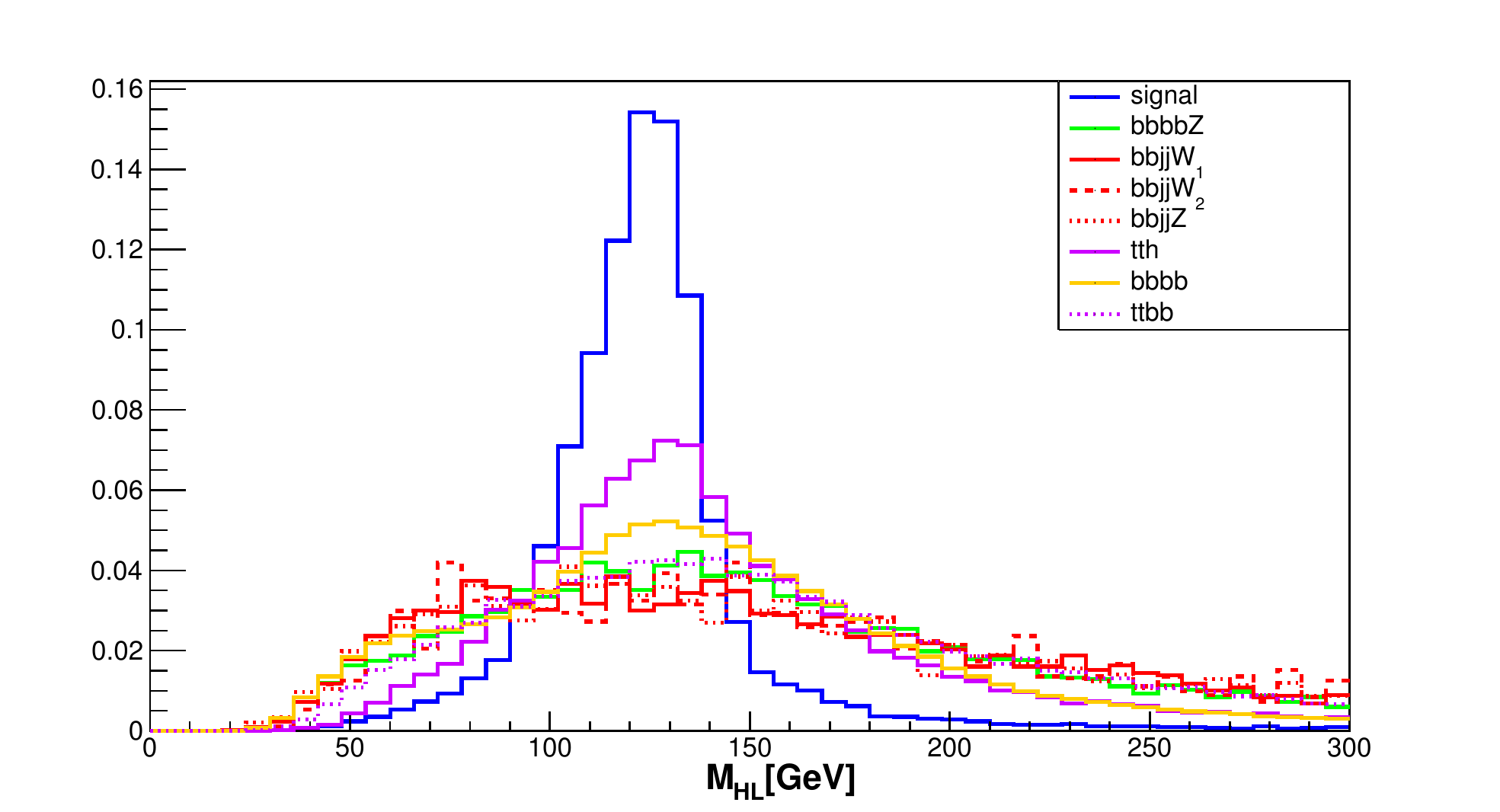}
     \includegraphics[width=7.7cm]{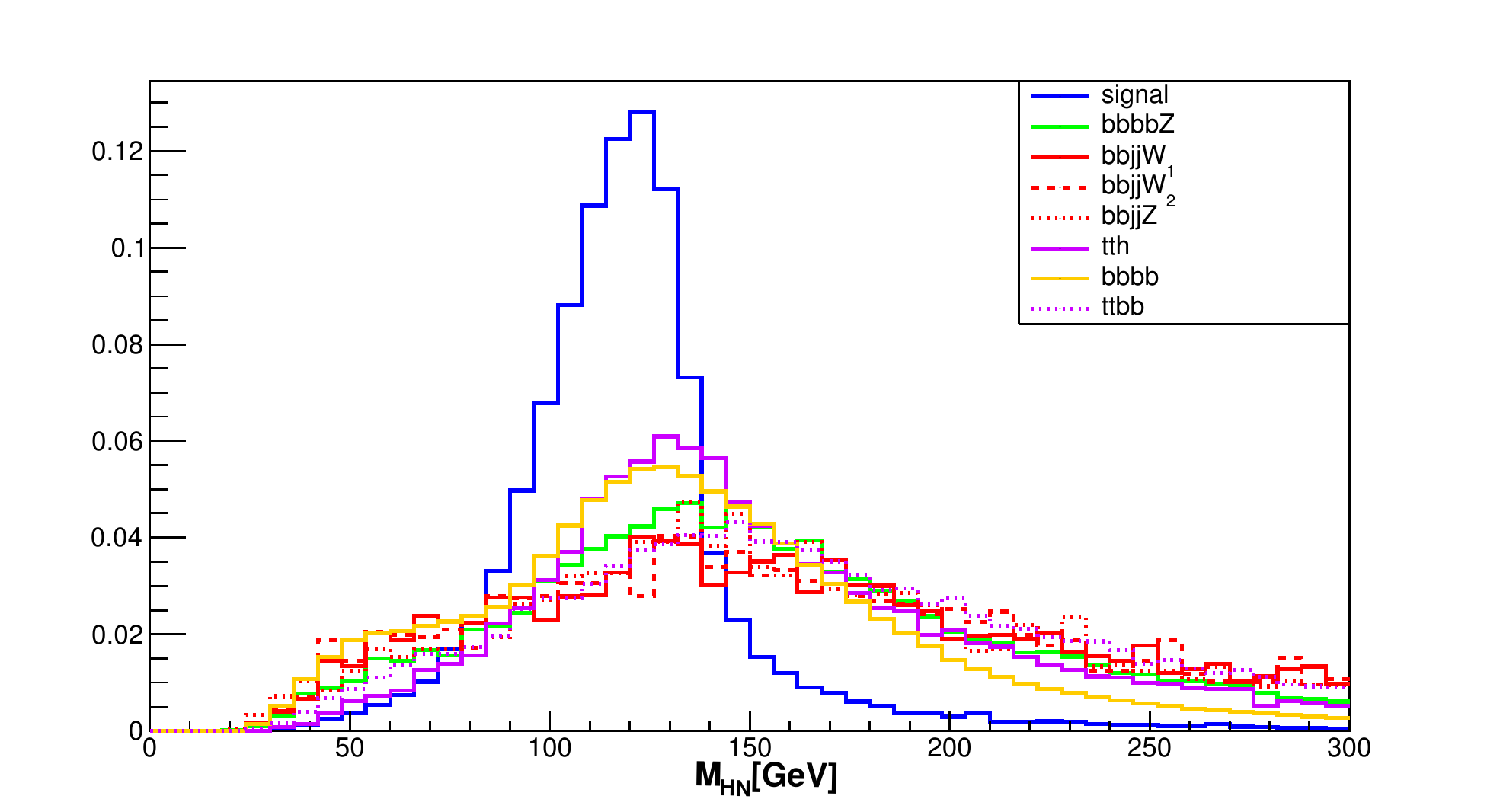}
     \caption{\label{fig:var1b2}
     Distributions of signal and backgrounts after basic cuts for $\mu=150$ GeV. The $y$-axis represents the normalized number of events}
     
     \hspace{-0.5cm}
 \end{figure}

    The distributions of the signal and backgrounds are displayed in Fig.~\ref{fig:var1b2} for $\mu = 150$ GeV. 
      It is easy to find that the distributions of the invariant masses of the leading and next-to-leading Higgs bosons ($M_{HL}$ and $M_{HN}$) and the corresponding azimuths ($\DeltaR_{HL}$ and $\DeltaR_{HN}$) could be used to effectively suppress the background.
      The cuts 120 GeV $< M_{H(N,L)} <$ 130 GeV and $\DeltaR_{H(N,L)}< 2$ are used for the two reconstructed Higgs bosons. Note that the rather strong invariant mass cuts are applied since the $bbbb$ background is large. Here the missing energy $\met$ cannot suppress the QCD background very well, and thus we impose a relatively strong cut on missing energy $\met >$ 50 GeV to get a larger value of $S/B$. With all above cuts, the significance can only reach $0.679\sigma$, as shown in Table \ref{tab:Cutflow4b}.

\begin{table*}[t]
\caption{
Signal and background cross sections in fb after the cuts for $\mu$ = 150 GeV.  
The significance $\sigma$ is calculated for a luminosity of 3~$\rm{ab}^{-1}$ at the 14 TeV LHC.
} 
\label{tab:Cutflow4b}
\vspace*{-0.2cm}
\begin{center}
\setlength{\tabcolsep}{1.1mm}
\renewcommand{\arraystretch}{1.4}
\scalebox{0.77}{
\begin{tabular}{|c||c|c|c|c|c|c|c|c|c|c|}
\hline   
 Cuts & signal   
& $bbbbZ$   & $bbjjW_1$  & $bbjjW_2$  & $bbjjZ$  
& $tth(b\bar{b})$ & $b\bar{b}b\bar{b}$ & $t\bar{t}b\bar{b}$
& $\sigma$  & $S/B$ 
\\
\hline \hline 
130 GeV > $M_{H(L,N)}$ > 120 GeV
& 0.036 & 0.0014 & 0.005 
& 0.003 & 0.005 & 0.057 & 22.41 & 0.086
& 0.53 & 0.0016
 \\
\hline
 $\DeltaR_{H(L,N)}$ < 2
& 0.017 & 0 & 0.001 
& 0.0005 & 0 & 0.028 & 3.58 & 0.04
& 0.637 & 0.005
\\
\hline
$\met$ > 50 GeV
& 0.0073 & 0 & 0.001 
& 0 & 0 & 0.028 & 0.54 & 0.02
& 0.679 & 0.012
\\
\hline

\end{tabular}}
\end{center}
\end{table*}
 With the increase of $\mu$, the cross section of the signal decreases rapidly, and at the same time the missing energy and $P_T^H$ become larger, as shown in Fig.~\ref{fig:var1b1}. Now we take $\mu$ = 300 GeV as an example to show the analysis for a larger $\mu$. 
 The cuts 105 GeV <$M_{H(L,N)}$ <140 GeV, which keep most of the signal events, are used for paired b jets.  
 Then, since there are no obvious differences on angular separations between signal and backgrounds, we use the same cuts  $\DeltaR_{H(N,L)} < 2$ as before.  
 Unlike the case of $\mu$ = 150 GeV, in this case the missing energy $\met$ could play an important role in distinguishing signal and backgrounds. We set $\met > 190$ GeV and this cut can increase the significance by about a factor of ten. The detailed results could be found in Table \ref{tab:Cutflow4b2}.

\begin{table*}[t]
\caption{
Signal and background cross sections in fb after cuts at different stages for $\mu$ = 300 GeV.
} 
\label{tab:Cutflow4b2}
\vspace*{-0.2cm}
\begin{center}
\setlength{\tabcolsep}{1.1mm}
\renewcommand{\arraystretch}{1.4}
\scalebox{0.77}{
\begin{tabular}{|c||c|c|c|c|c|c|c|c|c|c|}
\hline   
 Cuts & signal   
& $bbbbZ$   & $bbjjW_1$  & $bbjjW_2$  & $bbjjZ$  
& $tth(b\bar{b})$ & $b\bar{b}b\bar{b}$ & $t\bar{t}b\bar{b}$
& $\sigma$  & $S/B$ 
\\
\hline \hline 
140 GeV > $M_{H(L,N)}$ > 105 GeV
& 0.039 & 0.018 & 0.044 
& 0.024 & 0.054 & 0.6 & 236.442 & 0.75
& 0.138 & 0.00013
 \\
\hline
 $\DeltaR_{H(L,N)}$ < 2
& 0.028 & 0.0047 & 0.014 
& 0.007 & 0.014 & 0.31 & 47.53 & 0.39
& 0.21 & 0.00058
\\
\hline
$\met$ > 190 GeV
& 0.0079 & 0.0005 & 0.001 
& 0.0005 & 0.0027 & 0.011 & 0.0173 & 0.011
& 2.02 & 0.18
\\
\hline

\end{tabular}}
\end{center}
\end{table*}
From the above results we find that the $\met$ cut plays a major role in distinguishing the signal from the backgrounds for a large $\mu$. The detailed cuts are summarized as follows: 

\begin{itemize}
    \item The four leading jets must be $b-$tagged with $P_{T}$ > 40 GeV;
    \item Exactly two Higgs are reconstructed, with  140 GeV >$M_{H(L,N)}$ > 105 GeV;
    \item Proximity cut of $\DeltaR_{H(N,L)}$ < 2 for the two Higgs;
    
    \item $\met > 190$ GeV for the reconstructed missing transverse momentum.
\end{itemize}

\begin{figure}[t]
     \centering
     \includegraphics[width=11.5cm]{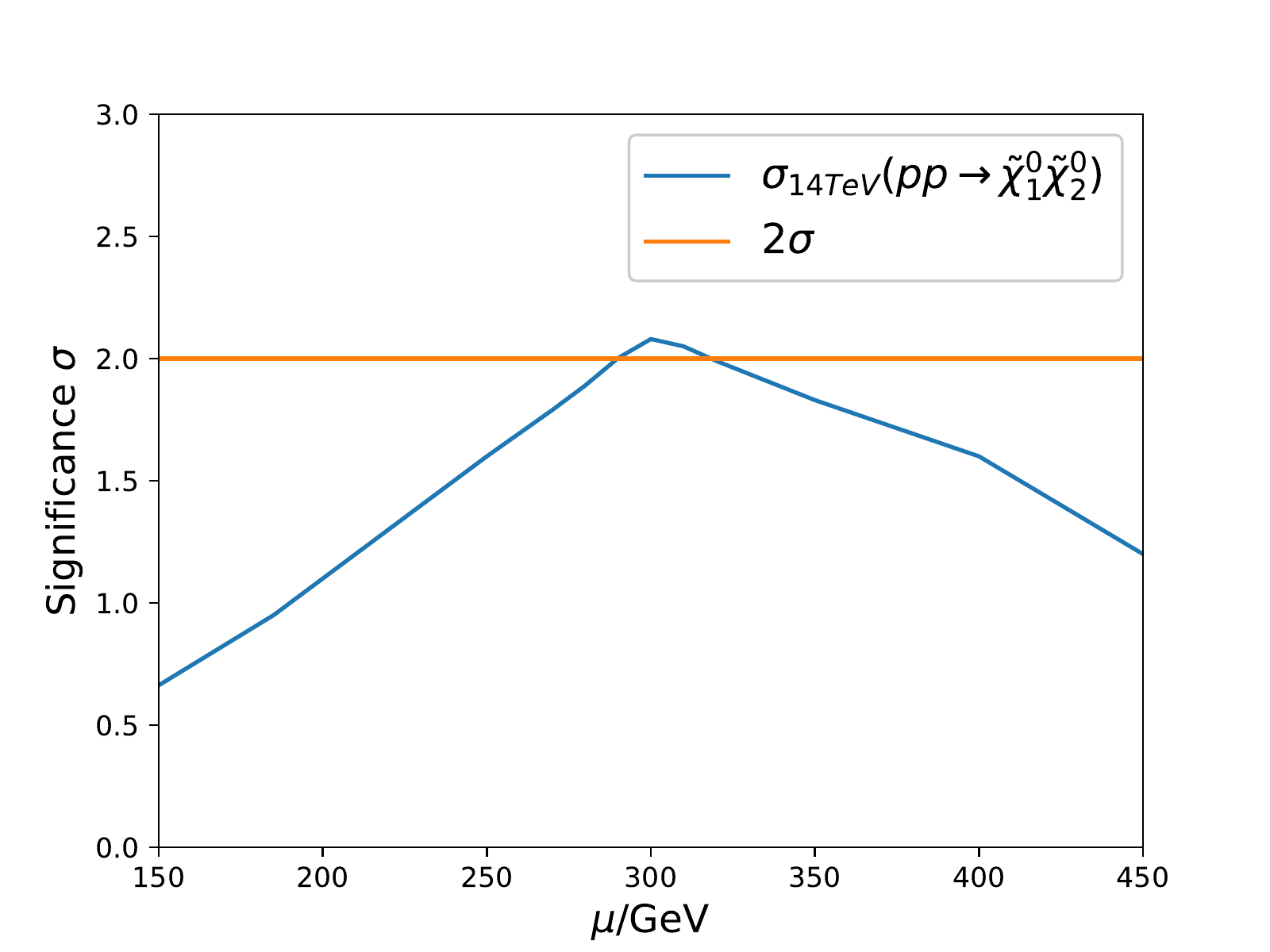}
     \caption{\label{fig:var1sig}
     The signal  $bb\bar{b}\bar{b}+\met$ significance versus the $\mu$ value at the HL-LHC (3 ab$^{-1}$, 14 TeV). }
     \hspace{-0.5cm}
 \end{figure}

 After the sequential cuts, we find that the multi-jets background are greatly suppressed, while the $t\bar{t}h,b\bar{b}b\bar{b},t\bar{t}b\bar{b}$ background remains large and  will be further cut with a larger $\met$. With the increase of $\mu$ value, the signal cross-section will be reduced but the cut with a larger $\met$ can help for the signal significance. For $\mu = 300$ GeV, the signal significance can reach to 2.02$\sigma$, as shown in Fig.~\ref{fig:var1sig}. As $\mu$ increases above 320 GeV the significance drops below 2$\sigma$. Note that in this scenario the $\mu$ value determines the lightest neutralino mass. 
 
%
\subsection{The signal of $hhG'G' \rightarrow b\bar{b}\gamma\gamma+\met$}
Finally we turn to the third channel with $h \rightarrow b\bar{b}$ and $h \rightarrow \gamma \gamma$.
It has a pair of photons reconstructed at the invariant mass around the Higgs boson. So the backgrounds include the single Higgs production such as $t\bar{t}H$. For non-resonant backgrounds and jet-fake backgrounds, we should consider missing energy based on the backgrounds considered in the preceding section. From the calculation we mainly need to consider the backgounds from $bbj\gamma W^{\pm}(W^{\pm}\rightarrow\ell^{\pm}\nu)$, $bbj\gamma W^{\pm}(W^{\pm}\rightarrow \tau^{\pm} \nu)$, $bbj\gamma Z(Z\rightarrow \nu \bar{\nu})$, $bbbbZ$, and $t\bar{t}\gamma$ productions.

The parton-level events for backgrounds are generated with MadGraph5, using the PDF CTEQ6L1~\cite{2003JPP}. For the signal calculation, it involves the loop-induced Higgs decay to di-photon and we use the loop-induced model~\cite{Hirschi:2015iia}. The parton-level cuts are imposed in order to avoid any divergence in the parton-level calculation~\cite{ATLAS:2017muo}: $P_{Tj} > 20$ GeV, $P_{T\gamma} > 25$ GeV, $|\eta_j|< 2.5$, $|\eta_{\gamma}| < 2.5$, $m_{jj}>25$ GeV.

\begin{figure}[t]
     \centering
     \includegraphics[width=7.5cm]{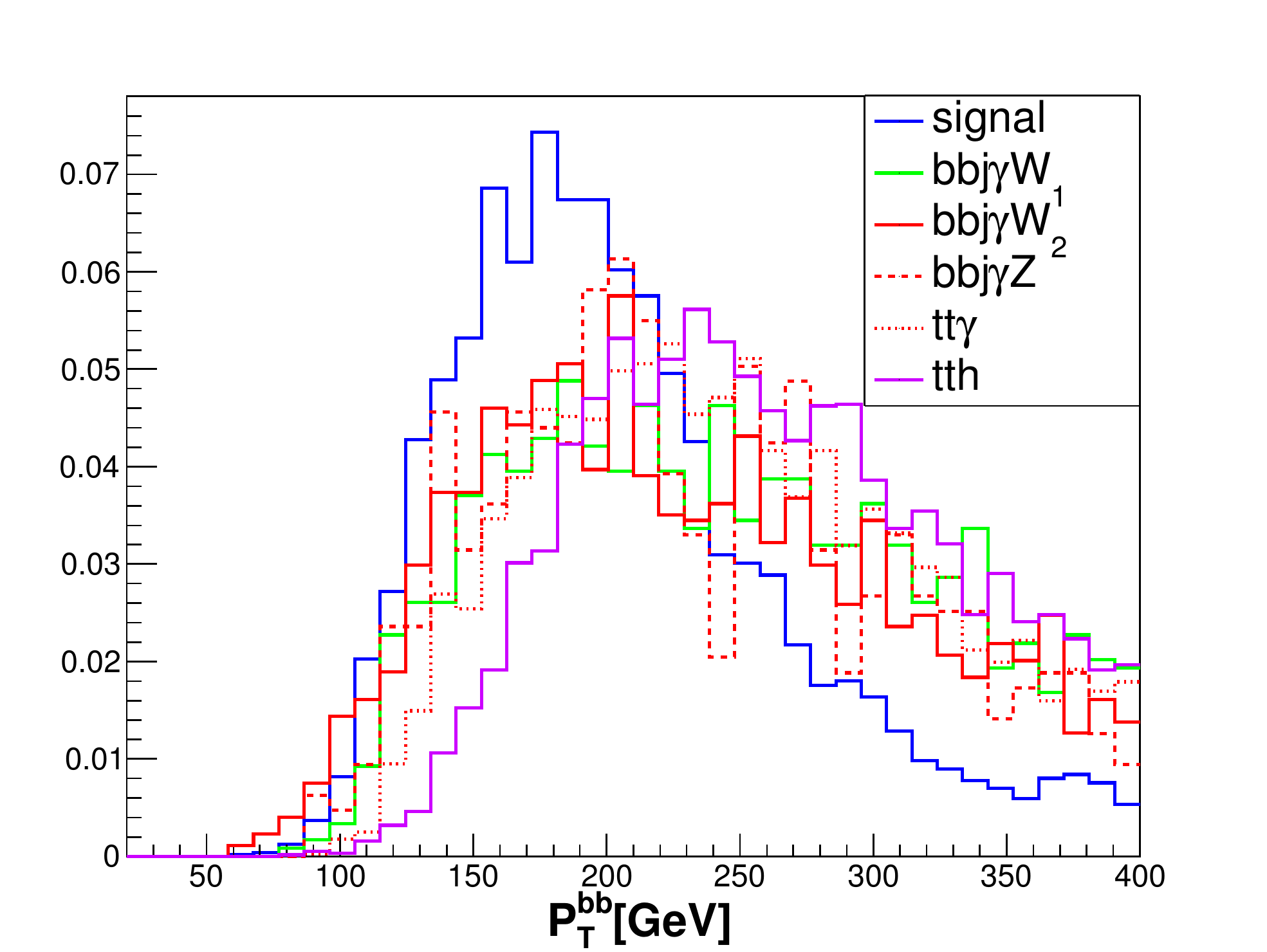}
     \includegraphics[width=7.5cm]{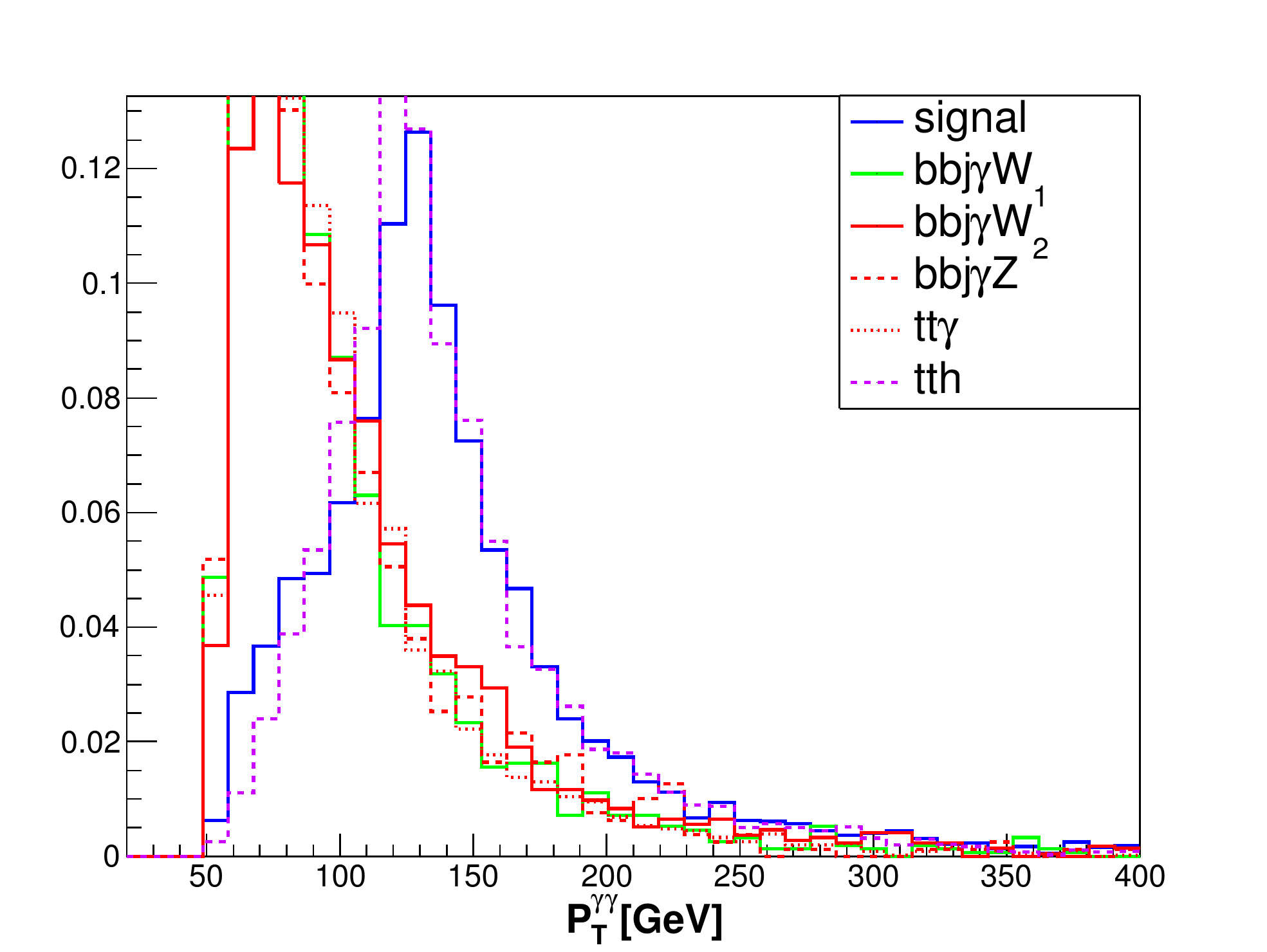}
     \\ \vspace*{-0.1cm}
    \includegraphics[width=7.5cm]{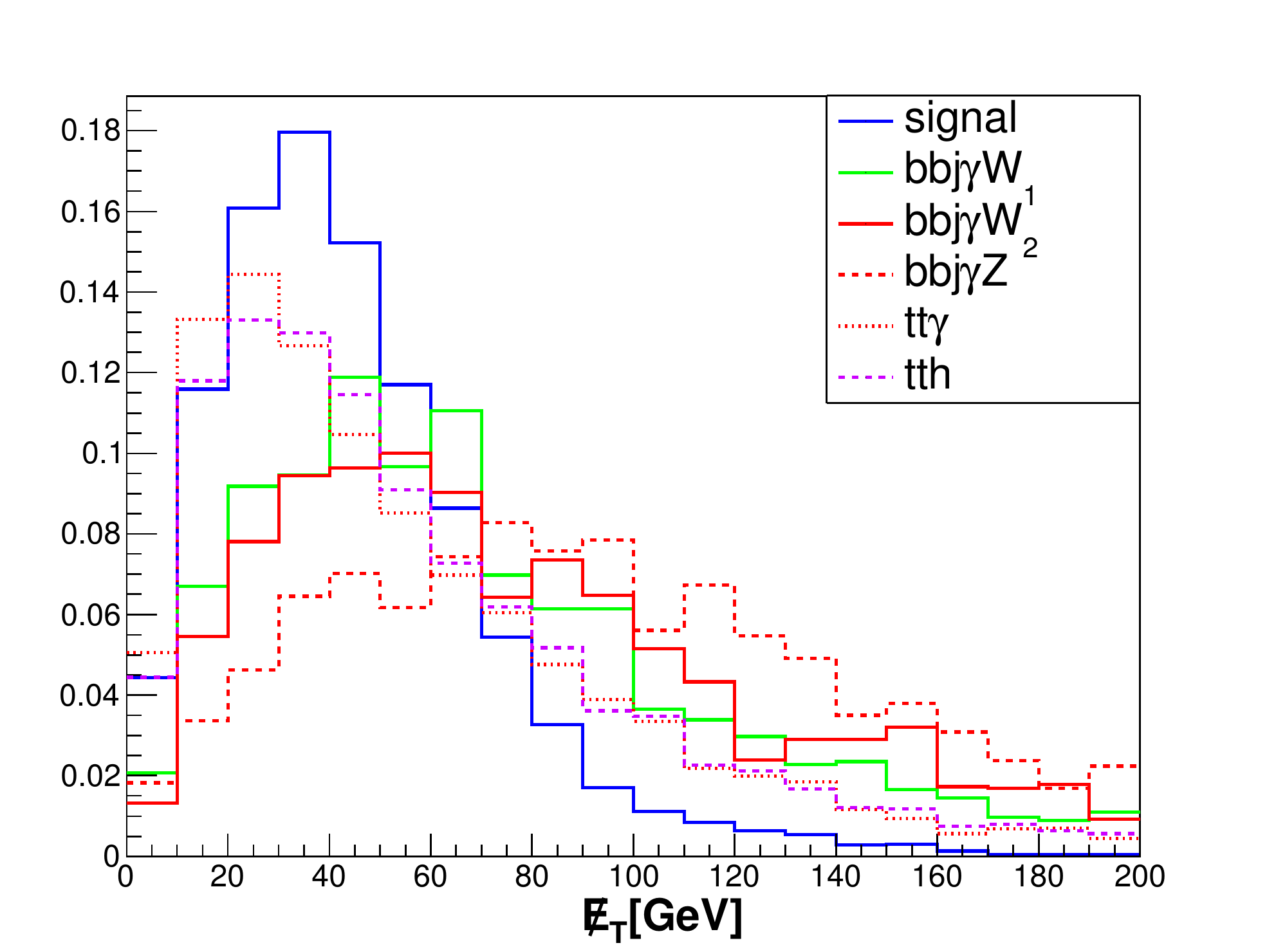}
     \includegraphics[width=7.5cm]{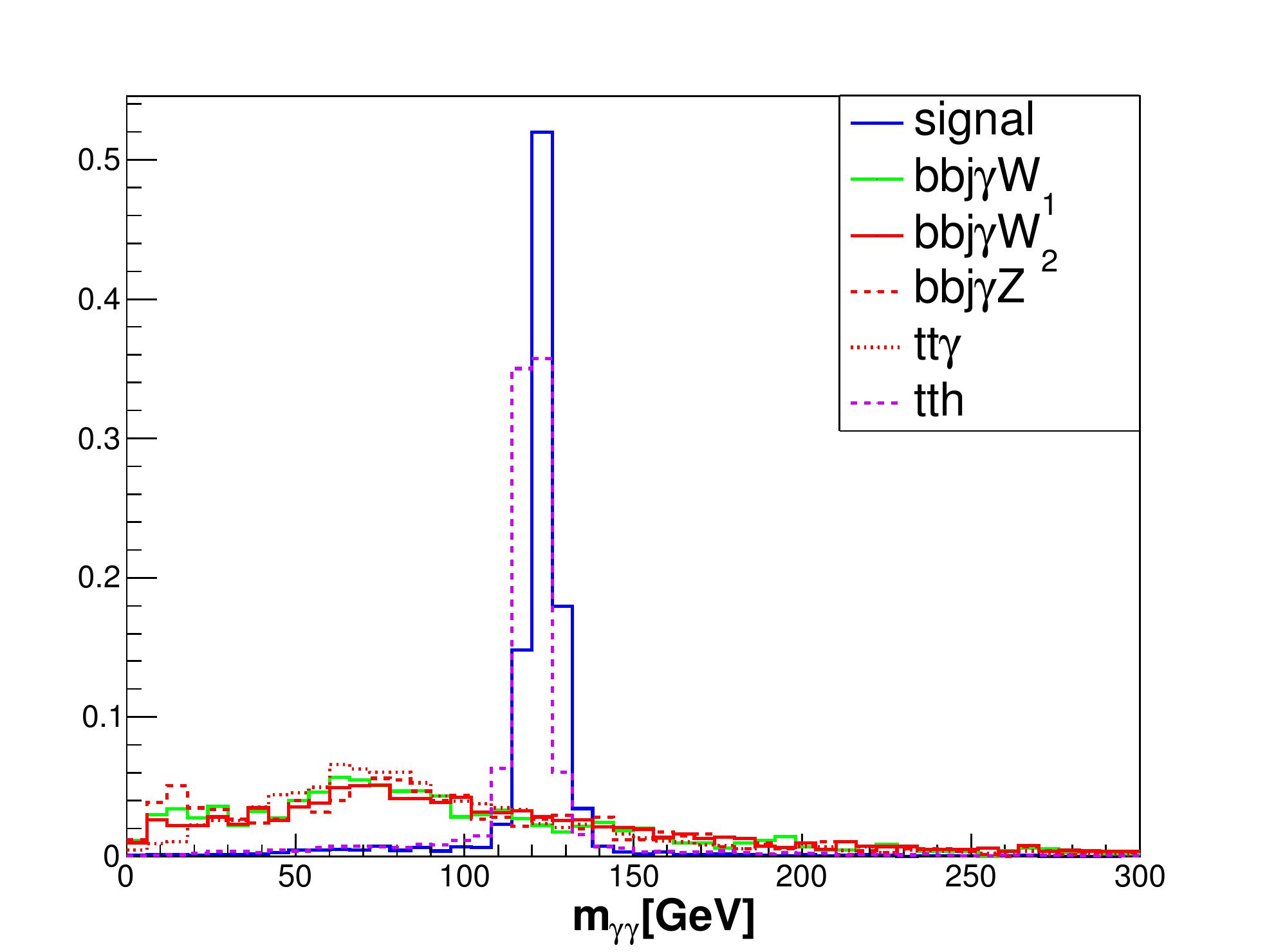}
     \\ \vspace*{-0.1cm}
     \hspace{-7.5cm} \includegraphics[width=7.5cm]{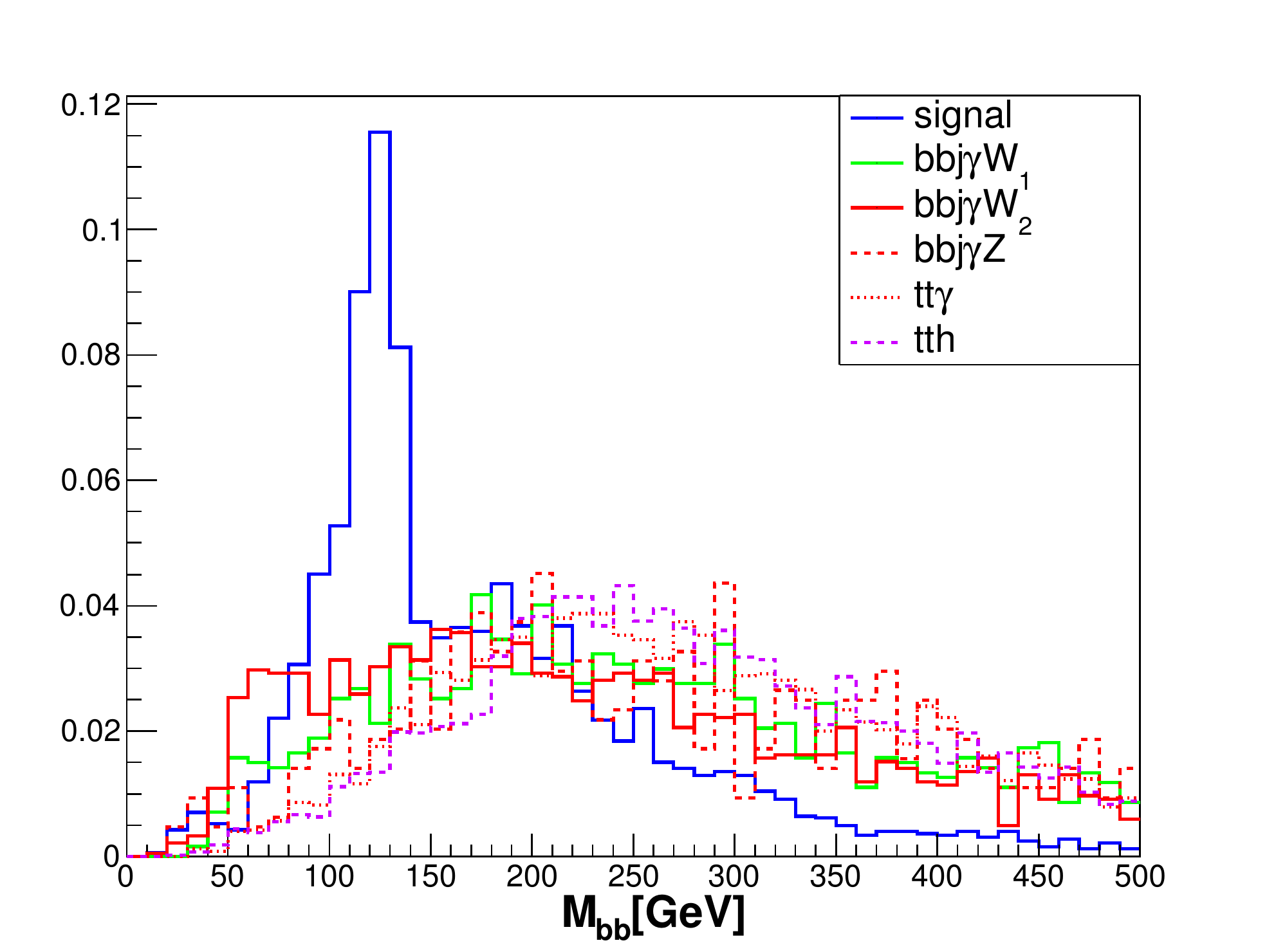}
     \caption{\label{fig:var1}
     Distributions for the signal with $\mu$ =150 GeV and backgrounts. The $y$-axis represents the normalized number of events for each process. }
     \hspace{-0.5cm}
 \end{figure}
For signal and backgrounds we require a pair of isolated photons and a pair of isolated $b$-tagged jets which are reconstructed near the Higgs boson mass. In Fig.~\ref{fig:var1}, we show the distributions of $P_T^{bb}$, $P_T^{\gamma\gamma}$, $\met$, $m_{\gamma\gamma}$ and $m_{bb}$ for signal and backgrounds. Since the angular separations have no obvious difference between signal and backgrounds, we set loose cuts $1.0 < \DeltaR_{bb} < 3.8$ and $1 < \DeltaR_{\gamma\gamma} < 3.7$. Obviously, $m_{\gamma\gamma}$ and $P_T^{\gamma\gamma}$ have distinctive features to distinguish signal from backgrounds expect $t\bar{t}h$. In order to reduce the background of $t\bar{t}h$, we require $P_T^{bb}$ in the range of $[80, 250]$ GeV and $P_T^{\gamma\gamma} > 70$ GeV. The $\met$ cut can efficiently suppress the multi-jet backgrounds and we set 10 GeV $< \met < 90$ GeV. The cuts can be summarized in the following:
\begin{itemize}
    \item The two leading jets must be $b-$tagged with 250 GeV $> p_T^{bb} > 80$ GeV;
    \item Exactly two isolated photons with 
    $ p_T^{\gamma\gamma} > 70$ GeV;
    \item Proximity cut of $1 < \DeltaR{_{\gamma\gamma}}$<3.7 for the two photons;
    \item Proximaty cut of $1 < \DeltaR{_{bb}} < 3.8$ for the two $b-$tagged jets;
    \item 121 GeV $< m_{\gamma \gamma} < 128$ GeV for the two photons;
    \item 100 GeV $< m_{bb} < 150$ GeV for the two $b-$tagged jets;
    \item 90 GeV $> \met =|\mptvec|> 25$ GeV for the reconstructed missing transverse momentum.
\end{itemize}

The detailed results are shown in  Table \ref{tab:Cutflowba}.
From our analysis we find that the $t\bar{t}\gamma$ background still remains large after these cuts. The signal significance can only reach to $0.79\sigma$. 

%
\begin{table*}[t]
\begin{center}
\setlength{\tabcolsep}{1.1mm}
\renewcommand{\arraystretch}{1.4}
\caption{
The cut flow for signal and background cross sections in fb for a luminosity of 3~$\rm{ab}^{-1}$ at the 14 TeV LHC.
} 
\label{tab:Cutflowba}
\vspace{.5cm}

\scalebox{0.77}{
\begin{tabular}{|c||c|c|c|c|c|c|c|c|}
\hline   
 Cuts & signal   
& $b\bar{b}j\gamma W_1$   & $b\bar{b}j\gamma W_2$  & $b\bar{b}jZ$  & $t\bar{t}\gamma$  
& $t\bar{t}h$ 
& $\sigma$  & $S/B$
\\
\hline \hline 
80 GeV < $P_T^{bb}$ < 250 GeV, $P_T^{\gamma\gamma}$ >70 GeV 
& 0.0093 & 0.0466 & 0.036
& 0.044 & 4.09 & 0.011
& 0.25 & 0.0022
 \\
\hline
100 GeV < $m_{bb}$ < 150 GeV, GeV 121 $< m_{\gamma\gamma}$ < 128 GeV   & 0.0051 & 0.0008 & 0.001 & 0.0005 & 0.089 & 0.0026
& 0.9 & 0.054
\\
\hline
90 GeV > $\met$ > 25 GeV   & 0.0035 & 0.0006  & 0.0009 & 0 & 0.052 & 0.002
& 0.79 & 0.62
\\
\hline
\end{tabular}}
\end{center}
\end{table*}

Some remarks are in order. From the above results we know that for the three channels only the $bbbb+\met$ signal might be observable at the HL-LHC. Note that some machine learning methods~\cite{Blanke:2019hpe,Amacker:2020bmn,Abdughani:2019wuv,Abdughani:2020xfo} which have been used for many body final states, can help to enhance the signal significance. Since our main aim is to show the new signature that comes up with pseudo-goldstino and to study the role of $\met$ in distinguishing the signal from backgound, only the traditional cut-flow Monte Carlo simulations were used in our analysis.
In addition, we did not consider the explanation of the muon $g-2$ 
in our scenario, which was found (see, e.g., \cite{Abdughani:2019wai, Athron:2021iuf, Wang:2021bcx, Endo:2021zal})
to require light neutralinos/charginos and light sleptons, possibly accessible at the HL-LHC \cite{Abdughani:2019wai}. 
For the status of low-energy supersymmetry confronted with  current experimental constraints, such as collider searches, dark matter searches and muon $g-2$ measurement, recent overviews could be found in Refs.~\cite{Baer:2020kwz,Wang:2022rfd}.    

Finally, we point out that our study may be also applicable to higgsino decay into a light singlino (which can be rather light as the lightest superparticle \cite{Cao:2013mqa}) plus a Higgs boson in the NMSSM \cite{Ellwanger:2009dp} which introduces a singlet Higgs superfield, mixing with the Higgs doublets, and thus relieves the 125 GeV Higgs mass constraint on stop masses \cite{Cao:2012fz}. 

\section{Conclusion}\label{sec:IV}
We considered the scenario of multi-sector SUSY breaking which predicts light pseudo-goldstinos 
and opens the decay mode of neutralino into pseudo-goldstino plus Higgs boson insider the detector at the LHC, leading to the signal of Higgs pair plus missing energy. We studied the observability of such Higgs pair plus missing energy from the decay of neutralino produced at the HL-LHC (14 TeV with a luminosity of 3~$\rm{ab}^{-1}$). 
We considered light higgsinos assumed in natural SUSY as an example and studied the three decay channels of the Higgs pair ($bbWW^*$, $bb\gamma\gamma$, $bbbb$). From detailed Monte Carlo simulations for the signal and backgrounds, the following observations are obtained: (i) The best channel is $bbbb+\met$, whose statistical significance can reach $2\sigma$ for a light higgsino in natural SUSY allowed by current experiments; (ii) The channels  $bb\gamma\gamma+\met$ and $bbWW^*+\met$ respectively give maximal statistical significance of $1\sigma$ and $0.79\sigma$. 
So the Higgs pair plus missing energy signal from the higgsino decay can be over the SM Higgs pair result which is about $1.8\sigma$ at the HL-LHC.

\section*{Acknowledgments}
This work was supported in part by IHEP under Grant No. Y9515570U1,
by the National Natural Science Foundation of China (NNSFC)
under grant Nos. 11821505 and 12075300, by Peng-Huan-Wu Theoretical Physics Innovation
Center (12047503), by the CAS Center for Excellence in Particle Physics (CCEPP), by the
CAS Key Research Program of Frontier Sciences, and by a Key R\&D Program of Ministry
of Science and Technology of China under number 2017YFA0402204, and by the Key Research Program of the Chinese Academy of Sciences, Grant NO. XDPB15.

\section*{Note added:}
When preparing this manuscript, the CMS collaboration reported 
their search results for the signal of Higgs pair plus missing energy using an integrated luminosity of 137 fb$^{-1}$ ~\cite{CMS:2022vpy}. 
When applied to higgsino decay into Higgs plus pseudo-goldstino, 
a range 305 GeV $>m_{\tilde{\chi}_1^0}(\simeq \mu) >$ 265 GeV can be excluded by this CMS search, assuming the decay branching ratio to be 1. Since in our study we assumed the branching ratio of higgsino decay to Higgs plus pseudo-goldstino to be 0.5, this excluded range 305 GeV $>m_{\tilde{\chi}_1^0}(\simeq \mu) >$ 265 GeV will become much narrower when applied to our scenario.


\bibliographystyle{JHEP}
\bibliography{draft}

\end{document}